\documentclass[usenames,dvipsnames,twocolumn]{aastex631}
\usepackage{fontenc}
\usepackage{graphicx}
\usepackage{dcolumn}
\usepackage{bm}
\usepackage{url}
\usepackage{tikz}
\usepackage{booktabs}
\usepackage{array}
\usepackage{multirow}
\usepackage{hyperref}
\usepackage{mathtools}
\usepackage{cancel}
\usepackage{subfigure}
\usepackage{comment}
\usepackage{float}
\usepackage[normalem]{ulem}
\usepackage{graphicx}
\usepackage{diagbox}
\usepackage{txfonts}

\begin{document}

\title{GRMHD study of accreting massive black hole binaries in astrophysical environment: a review}

\author{Federico Cattorini}
\correspondingauthor{Federico Cattorini}
\email{fcattorini@uninsubria.it}
 \affiliation{Dipartimento di Fisica G. Occhialini, Universit\`a di Milano-Bicocca, Piazza della Scienza 3, I-20126 Milano, Italy}
\affiliation{INFN, Sezione di Milano-Bicocca, Piazza della Scienza 3, I-20126 Milano, Italy}%

\author{Bruno Giacomazzo}
 \affiliation{Dipartimento di Fisica G. Occhialini, Universit\`a di Milano-Bicocca, Piazza della Scienza 3, I-20126 Milano, Italy}%
 \affiliation{INFN, Sezione di Milano-Bicocca, Piazza della Scienza 3, I-20126 Milano, Italy}%
 \affiliation{INAF, Osservatorio Astronomico di Brera, Via E. Bianchi 46, I-23807 Merate, Italy}%

\date{\today}

\begin{abstract}
We present an overview of recent numerical advances in the theoretical characterization of massive binary black hole (MBBH) mergers in astrophysical environments. These systems are among the loudest sources of gravitational waves (GWs) in the universe and particularly promising candidates for multimessenger astronomy.
Coincident detection of GWs and electromagnetic (EM) signals from merging MBBHs is at the frontier of contemporary astrophysics.
One major challenge in observational efforts searching for these systems is the scarcity of strong predictions for EM signals arising before, during, and after merger. 
Therefore, a great effort in theoretical work to-date has been to characterize EM counterparts emerging from MBBHs concurrently to the GW signal, aiming to determine distinctive observational features that will guide and assist EM observations. 
To produce sharp EM predictions of MBBH mergers it is key to model the binary inspiral down to coalescence in a full general relativistic fashion by solving Einstein's field equations coupled with the magnetohydrodynamics equations that govern the evolution of the accreting plasma in strong-gravity. We review the general relativistic numerical investigations that have explored the astrophysical manifestations of MBBH mergers in different environments and focused on predicting potentially observable smoking-gun EM signatures that accompany the gravitational signal. 
\end{abstract}


\section{Introduction}
The study of massive binary black hole (MBBH) evolution dates back as far as the early eighties. In their pioneering work, \citet*{Begelman-1980} explored the possibility that following a galaxy coalescence event, the active nucleus of the post-merger galaxy may harbor two massive black holes (MBHs) in orbit about each other.

Once two galaxies merge, the MBHs in their nuclei sink to the center of the merger remnant via a number of mechanisms such as dynamical friction via interaction with surrounding stars, individual interactions with singular stars, and torques exerted by a circumbinary gaseous disk \cite[see, e.g.,][for a review]{Dotti-2012}. Following these interactions, the binary will enter the so-called gravitational wave (GW) regime where it will evolve until merger. 
Gravitational waves are perturbations of the geometry of spacetime produced by the acceleration of massive objects \cite[][]{Dirkes-2018-GWREV}, which propagate away from the sources at the speed of light. Since the first detection of GWs by the Laser Interferometer Gravitational-Wave Observatory (LIGO)/Virgo collaboration in 2015 \cite[][]{Abbott-2016}, GW observations have opened a new way to observe and characterize binary systems of compact objects, such as (stellar-mass) black hole binaries \cite[][]{GWTC-2.1, GWTC-3}, neutron-star binaries \cite[][]{Abbott-2017}, and black hole-neutron star binaries \cite[][]{Abbott-2021}.

The binary evolution in the GW regime (Fig. \ref{fig:binary_phases}) is commonly divided into three phases, i.e. \textit{inspiral}, \textit{merger}, and \textit{ringdown} (for further knowledge on the features of propagating gravitational radiation and GW sources we defer the reader to \cite{Maggiore-Vol1}).
The classification into three (approximately defined) evolutionary stages is motivated by the different techniques used to describe each phase:
\begin{figure*}
\begin{center} 
\includegraphics[width=.9\textwidth]{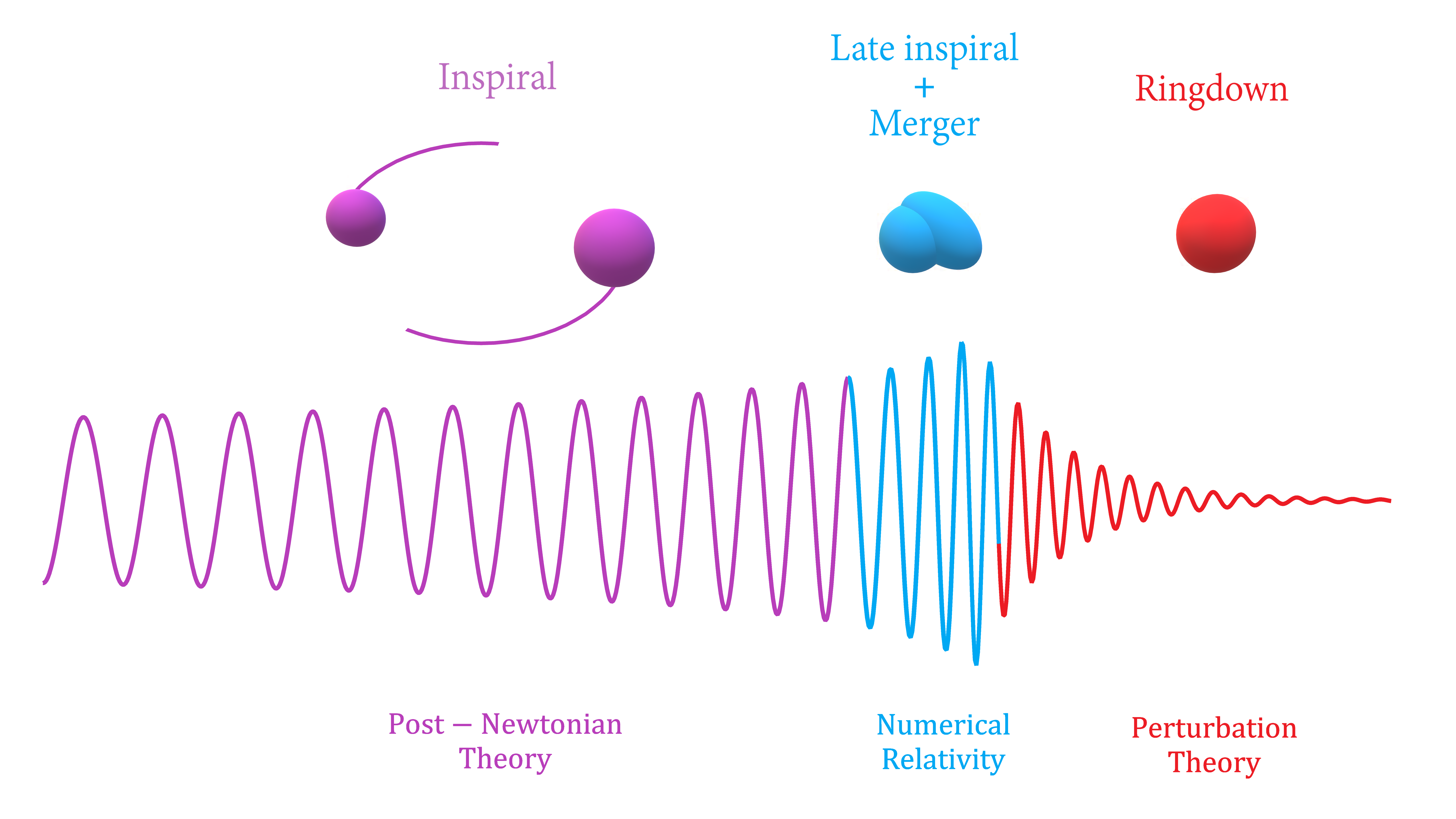}
\end{center}
\caption{The three phases in the temporal evolution of a binary system. In the inspiral phase (i), the black holes' inward radial migration timescale is much longer than their orbital timescale $t_{\mathrm{orb}}$. In the merger phase (ii), the two objects merge into one and the GW signal reaches its peak. In the ringdown phase (iii), the resulting object relaxes to a stationary Kerr configuration by emitting GWs at characteristic complex frequencies.}\label{fig:binary_phases}
\end{figure*}

\begin{itemize}
    \item[(i)] The inspiral is the most protracted stage of the GW-driven orbital evolution, in which the binary is well-separated and the black holes' inward radial migration timescale is much longer than their orbital timescale $t_{\mathrm{orb}}$. This stage can be treated analytically and is modelled using Post-Newtonian (PN) techniques.
    \item[(ii)] The second stage of the GW-driven binary evolution includes the late stages of the inspiral and the merger (or coalescence), in which the black holes plunge together and merge, while the GW emission reaches its peak. This phase (in which strong field, fully non-linear gravity plays a paramount role) must be treated numerically.
    \item[(iii)] The third phase is ringdown, in which the merger remnant settles down into a Kerr configuration by emitting GWs at characteristic complex frequencies. It can be modelled analytically with perturbation methods.
\end{itemize}
Exploring the second stage of the GW-driven binary evolution requires solving on a computer the full Einstein's equations of general relativity
\begin{equation}
    G_{\mu\nu}=8\pi T_{\mu\nu},
\end{equation}
in which we assumed the geometric unit system for which $c=G=1$.
This is a formidable task and took numerical relativists several decades and a great deal of trials and errors to overcome. In the past two decades, a number of remarkable breakthroughs has yielded robust and accurate full GR simulations of inspiralling and merging binary black holes, contributing to lay the foundations of gravitational wave astrophysics \cite[][]{Pretorius-2005, Baker-2006, Campanelli-2006, Campanelli-2006b, Berti-2007-unequalmass}. More recently, an increasing number of general relativistic magnetohydrodynamic (GRMHD) numerical studies have begun to explore the strong-field dynamics of the gas in the vicinity of massive binary black holes, investigating the mechanisms that might potentially drive electromagnetic (EM) emission concurrent to the gravitational radiation \cite[][]{Palenzuela-2010, Noble-2012, Farris-2010, Farris-2012, Gold-2014a, Gold-2014b, Giacomazzo-2012, Kelly-2017, Cattorini-2021}.

We direct the reader to \cite{Gold-2019} for a review of theoretical works on accreting black holes in the strong gravitational regime \cite[see also][]{Schnittman-2013a}. \cite{Gold-2019} presents an overview of the theoretical predictions on these systems, focusing on results from simulations of black hole binaries surrounded by a circumbinary accretion disk. Also, the recent review by \cite{Bogdanovic-2022-LR} provides a comprehensive introduction to the current body of knowledge about the nature of EM counterparts to MBBH mergers, both on the observational and theoretical sides.

The present work offers a complementary revision of the subject, presenting the current state-of-the-art of GRMHD simulations and discussing the diverse environments in which a massive binary may be embedded in the late stages of the GW-driven inspiral.

\section{Numerical relativity simulations of binary black holes}\label{sec:literature_rev}
Most numerical relativity simulations start with decomposing the four-dimensional spacetime into a set of non-intersecting three-dimensional spatial hypersurfaces parametrized by a time function $t$. In the 1960s,  \citet*{Arnowitt-1962} pioneered this “3+1" approach with their Hamiltonian formulation of GR. During the following years, a number of attempts to numerically evolve the head-on collision of two equal-mass black holes using the ADM formalism were carried out \cite[][]{Eppley-1975, Smarr-1975, Smarr-1976, Smarr-1977}. These simulations were two-dimensional and employed an axisymmetric approach, managing to evolve the systems to the collision and to extract information about the emitted gravitational signal. To this aim, they employed the most powerful computers in their day (as an example, the code presented in \cite{Smarr-1975} ran on the UT-CDC system, a mainframe calculator manufactured by Control Data Corporation with a 60 bit central processor performing at 10 MHz).
Unfortunately, the next steps \textendash \ three-dimensional simulations of orbiting binaries \textendash \  were not feasible at the time, due to insufficient computational resources and numerical instabilities, e.g., the “grid-stretching'' problem, which leads to divergent metric coefficients \cite[][]{Bruegmann-1999}. Consequently, the numerical exploration of merging binary black holes laid dormant for years.

\vspace{.2cm}
The tides turned during the late 1990s and into the early 2000s, when a number of important results were accomplished. These key developments include: new grid-based methods for representing black holes such as punctures \cite[][]{Brandt-1997a} and excision \cite[][]{Seidel-1992}; recognition of the importance of the hyperbolicity in formulating the Einstein's equations \cite[][]{Bona-Masso-1992, Abrahams-1997}; the improved BSSNOK formulation of the Einstein's equations \cite[][]{Nakamura-1987, Shibata-Nakamura-1995, Baumgarte-Shapiro-1998}; singularity-avoiding coordinate conditions \cite[][]{Bona-1995, Alcubierre-2003}; the Cactus computational framework \cite[][]{Goodale:2002a}, including several codes apt to the numerical evolution of general relativistic systems.
\begin{figure*}
\begin{center} 
\includegraphics[width=\textwidth]{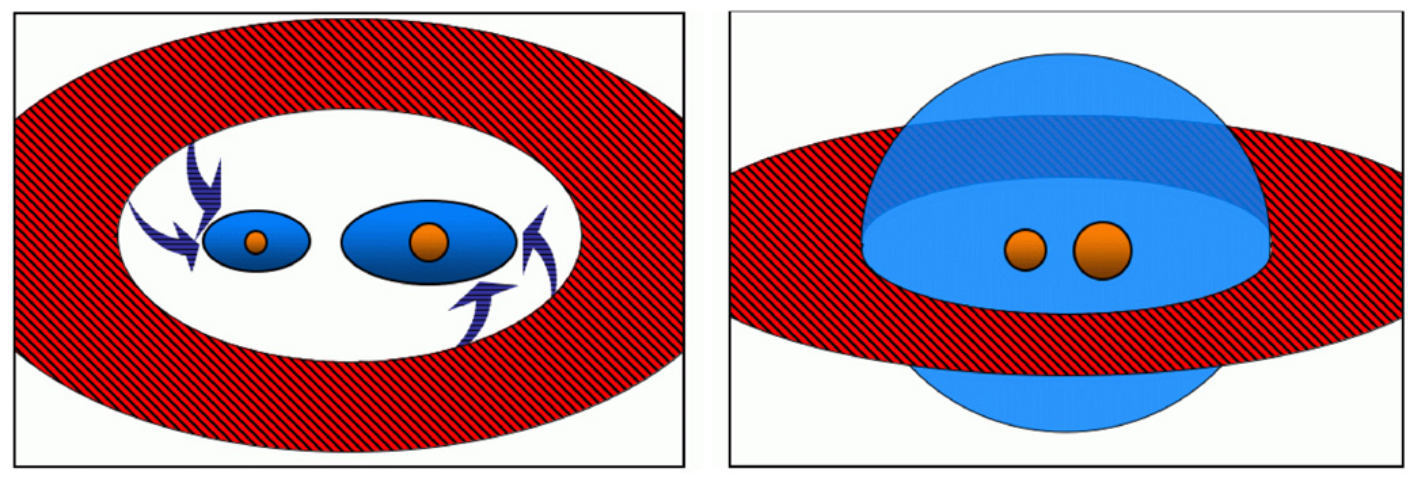}
\end{center}
\caption{Artistic illustration of the two accretion scenarios discussed in this section. Left: circumbinary disk model in which binary torques create a central low-density cavity. The accretion onto the individual black holes is mediated by individual mini-disks around each BH. Right: gas cloud scenario in which MBBHs remain engulfed in the hot thin gas until coalescence. At larger radii, the geometrically thick accretion flow transitions into a cold, geometrically thin disk. The illustrations are not to scale. Figure adapted from \cite{Bogdanovic-2011}.}\label{fig:MBBH_scenarios}
\end{figure*}

Eventually, in 2005, the pioneering work by \cite{Pretorius-2005} ushered in the first stable numerical evolution of merging BBHs in vacuum. Later in 2005, the groups at University of Texas at Brownville (UTB) and NASA's Goddard Space Flight Center independently developed a new technique, called “moving punctures", that also produced successful simulations of binary black hole mergers \cite[][]{Baker-2006, Campanelli-2006}. Since then, outstanding progress has been made to explore numerically the late stages of binary evolution and it is now possible to perform stable and accurate general relativistic simulations of BBHs in vacuum exploring a wide range of parameter space \cite[][]{Healy-2022_4thRITcat}, including mass ratios\footnote{The mass ratio is defined as $q \equiv m_2/m_1$, where $m_1$ ($m_2$) is the mass of the primary (secondary) BH.} as small as $q= 1/128$ \cite[][]{Lousto-2020PhRvL.125s1102L}, and nearly extremal spin parameters $\chi = S_i/m_i^2 = 0.99$ \cite[][]{Lovelace-2010, Zlochower-2017PhRvD..96d4002Z}.

After the initial breakthrough with equal-mass, non-spinning black holes, the remarkably robust “moving puncture” method was soon applied to a wide variety of systems, including unequal masses \cite[][]{Berti-2007-unequalmass}, eccentric orbits \cite[][]{Hinder-2008-eccentric}, and spinning BHs \cite[][]{Campanelli-2006b}, and generic black hole binaries, i.e., binaries containing unequal-mass black holes with misaligned spins \cite[][]{Campanelli-2007}.

\section{Magnetohydrodynamic simulations of MBBHs}\label{sec:lit_rev_matter}
While numerical relativity had initially been devoted to applications to the direct detection of GWs, much of the recent theoretical work has focused on predicting potentially observable electromagnetic (EM) signals emerging from MBBHs.
In fact, since galaxy mergers are usually followed by gas inflows toward the center of the merger remnant, the gravitational signal emitted by MBBH mergers is expected to be accompanied by an EM counterpart generated by the gas accreting onto the merging binary \cite[][]{Bogdanovic-2022-LR, LISA2022}.

One step toward the general relativistic modelling of the accretion flows onto merging MBBHs was taken by \cite{vanMeter-2010}. This work produced three-dimensional general relativistic simulations of merging binary black holes and modelled the environment around the binary as a flow of pressureless matter made up by non-interacting point particles (i.e., particles of non-zero rest mass in geodesic motion) evolved with particle tracing techniques. While pressureless-matter simulations can be computationally efficient and offer certain insights, they lack the physical depth required to explore realistic accretion flows, which involve the interaction of gas, radiation, and magnetic fields. The exploratory simulations by \cite{vanMeter-2010} lacked the required sophistication needed to model the gaseous environment in the vicinity of merging binaries. In actual accretion flows onto black holes, gas often experiences compression and heating as it spirals inward, leading to the formation of shocks. Hydrodynamic (HD) simulations excel in capturing shock physics, including the dissipation of kinetic energy into thermal energy. Conversely, pressureless-matter simulations are unable to accurately model these shock formations and the associated energy dissipation processes. Furthermore, the realistic portrayal of accretion flows necessitates the consideration of the interaction between gas and magnetic fields.This intricate coupling can be effectively addressed through magnetohydrodynamic (MHD) calculations.

At present, our understanding of the features of these accretion flows is in its infancy. The structure of the accretion flows around coalescing MBBHs largely depends on the angular momentum content of the accreting gas conveyed in the galactic merger and on its thermodynamical state. Two limiting physically motivated scenarios bracket the range of properties of accreting fluids around MBBHs, i.e. the \textit{circumbinary disk} and the \textit{gas cloud} models (Fig. \ref{fig:MBBH_scenarios}). In the following we outline the properties of both scenarios and describe the results from recent state-of-the-art, general relativistic numerical simulations.

In the following, we will often express distances in units of the total mass of the system $M$. In GRMHD simulations (as long as that the total mass of the fluid is negligible with respect to the mass of the binary), one can neglect source terms in the spacetime evolution equations. If this is the case, the total mass of the system $M$ sets\footnote{As an example, the unit length for a binary system of total mass 1M$_{\odot}$ is $\hat{l}\approx 1.5$ km $M_0$ and the unit time is $\hat{t}\approx 5 \ \mu s \ M_0$, where $M_0\equiv M/1\mathrm{M}{\odot}$.} the length unit $\hat{l}=GM/c^2$ and the time unit $\hat{t}=GM/c^3$. In geometric units, $\hat{l}=\hat{t}=M$.
\subsection{Circumbinary disk scenario}\label{subsec:CBD}
Most theoretical studies in the past decade have focused on the circumbinary disk (CBD) model. If, at the point of formation of the gravitationally bound binary, the two MBHs have a mass ratio larger than a few percents, theoretical studies indicate that the binary torques can clear a central low density cavity with a radius corresponding to approximately twice the binary semimajor axis \citep[see, e.g.,][]{D'Orazio-2013, D'Orazio-2016}. 
Simulations also indicate that despite strong binary torques, accretion into the central cavity continues unhindered relative to the single MBH case \citep[e.g.,][]{D'Orazio-2013, Farris-2014a, Shi-2015}.

Circumbinary black hole accretion simulations have been conducted adopting diverse methods \cite[see][for a recent review of circumbinary accretion simulations incorporating relativistic effects]{Gold-2019}. In the following, we will focus on numerical relativity simulations that solve the GRMHD equations over dynamical spacetimes that are evolved either by solving the full set of Einstein's equations or with approximate techniques such as the so called Post-Newtonian (PN) gravity, which describes the spacetime by a high-order PN approximation.
One advantage of evolving three-dimensional GRMHD simulations with PN gravity is that they do not carry the burden of evolving the spacetime by solving the full set of Einstein's equations. Therefore, a larger amount of computational resources may be dedicated to the MHD study of the CBD regions \cite[][]{Noble-2012, Bowen-2018, Noble-2021}, including more realistic physical assumptions and covering sufficiently wide time scales in order to account for accurate variability analysis.

Conversely, a full general relativistic treatment of the spacetime evolution \cite[][]{Farris-2012, Gold-2014a, Gold-2014b, Giacomazzo-2012, Paschalidis-2021} is crucial to resolve the physics in the proximity of the BH horizons (e.g., the launching of jets from the interaction of magnetic fields with the horizons).
\subsubsection*{Post-Newtonian MHD simulations}\label{subsec:PN_RIT}
In the last decade, a variety of numerical groups have developed a number of high-end MHD simulations of circumbinary accretion onto MBBHs employing diverse successful approximations for the spacetime metric evolution.
The pioneering investigation by \cite{Noble-2012} presented a simulation of circumbinary accretion onto an equal-mass binary, evolving the GRMHD equations  with the \texttt{Harm3D} code \cite[][]{Noble-2009a} over a changing spacetime described by a high order \textit{Post-Newtonian (PN) approximation}.
This simulation provided a firm step toward the general relativistic investigation of a CBD surrounding a MBBH and established reasonable prior conditions for the gas that feeds these systems. To produce a quasisteady disk structure, \cite{Noble-2012} first ran a “secularly-evolving" configuration (RunSE) that kept the binary semimajor axis artificially fixed at a separation $R=20 M$, followed by an “inspiral" run (RunIn) that evolved the binary down to a separation of 8$M$. 

\begin{figure}
\includegraphics[width=.495\textwidth]{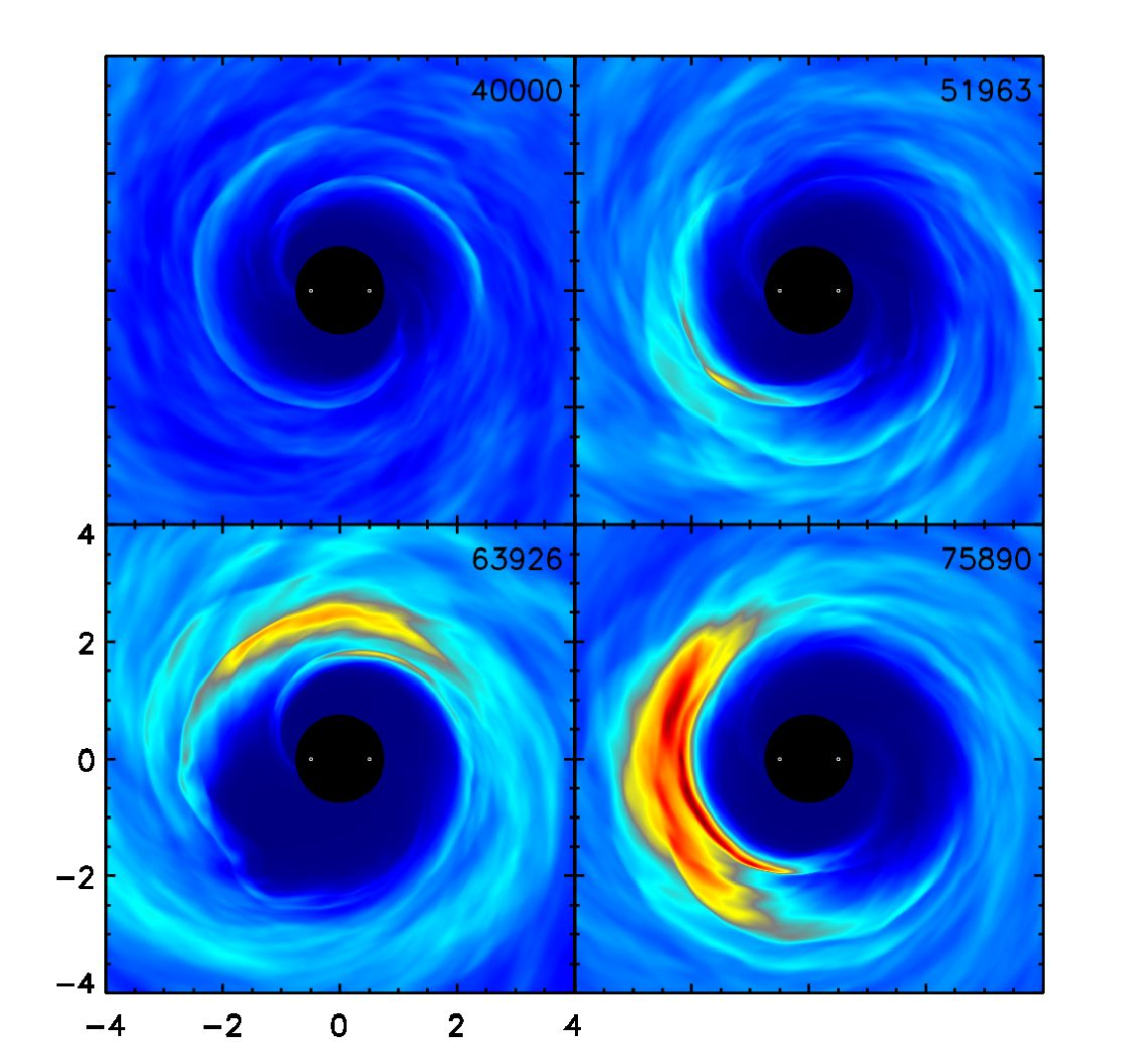}\\
\includegraphics[width=.495\textwidth]{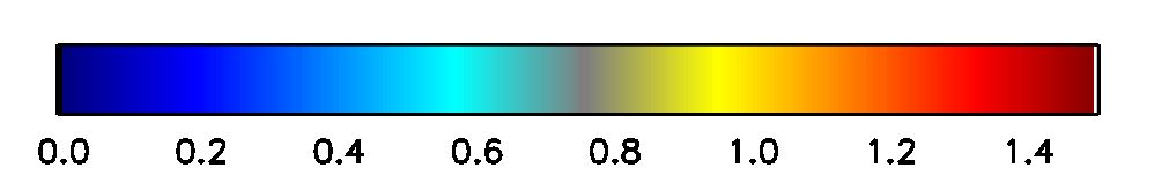}
\caption{Color contours of surface density in units of $\Sigma_0$ (i.e., the maximum surface density in the initial condition) as a function of radius and azimuthal
angle in RunSE at four different times. The linear color scale emphasizes the growth of the “lump" in the inner disk. The times shown are $t=40000M$ (upper-left), $t=51963M$ (upper-right), $t=63926M$ (lower-left), and $t=75890M$ (lower-right), where $M$ is the total gravitational mass of the system. Figure taken from \cite{Noble-2012}.}\label{fig:surfdens_equatorial}
\end{figure}

The RunSE evolution exploited the helical Killing symmetry \cite[][]{Klein-2004}, i.e. a particular symmetry of the spacetime that describes a binary held at fixed separation. This symmetry ensures that, while the separation is fixed, the spacetime is invariant in a frame corotating with the binary. Within such an asymmetric spacetime, the circumbinary disk is initialized by an axisymmetric gas distribution supported by pressure and rotation, and in hydrostatic equilibrium with a specific angular momentum profile \cite[see also][]{Chakrabarti-1985, Chakrabarti-1985-erratum, De_Villiers-2003}.
In their work, \cite{Noble-2012} followed the evolution of the gas surface density and found that torques driven by the binary can drive matter toward the inner binary domain (this central region was excised from the computations to avoid evolving the gas in the vicinity of the black holes).
Also, they found that a “lump" of matter \cite[][]{Shi-2012} forms in the disk, whose density contrast with respect to the adjacent regions grows steadily in time (see Fig. \ref{fig:surfdens_equatorial}). The subsequent investigation by \cite{Noble-2021} presented a suite of 3D GRMHD simulations with the same PN spacetime as model “RunSE'' by \cite{Noble-2012} and carried out a parameter exploration surveying different mass ratios and mass/magnetic fluxes in the accretion disks, aiming to linking the properties of the CBDs to the GW signal.

The circumbinary accretion and disk-feeding mechanisms were extensively explored by \cite{Bowen-2018, Bowen-2019}, which presented MHD simulation in which a CBD around a binary feeds mass to individual accretion disks (“mini-disks") around each MBH, and found that the streams of matter feeding the mini-disks comes into phase with the lump at the inner edge of the circumbinary disk. The simulation presented in these works evolve an equal-mass binary for approximately 12 orbital periods starting at an initial separation of 20$M$, thus allowing for a detailed examination of the asymmetric accretion flow and streams which deposit matter onto the mini-disks. The authors found that a number of aspects of the binary late inspiral can be characterized by quasiperiodic oscillations associated with the interaction of the lump with the individual MBHs. For instance, they identify quasiperiodic oscillations in the mass-flux of the accretion stream, passing (or “sloshing") from near one MBH to near the other and back again. 
\cite{Bowen-2018} suggested that the quasiperiodic modulations in the mini-disks structure may be related to characteristic time-dependent signatures in the EM emission of the binary.

\begin{figure}
\begin{center} 
\includegraphics[width=.5\textwidth]{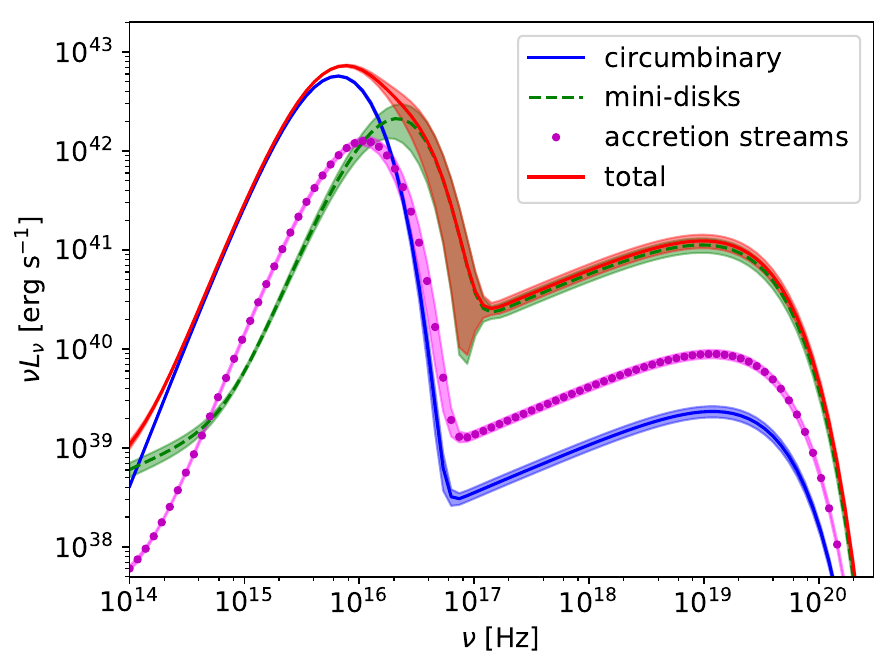}
\caption{Time-averaged luminosity ($\nu L_{\nu}$) spectrum obtained at $d=1000 M$ and $\theta=0$ using data from the second orbit of the binary simulation produced by \cite{Bowen-2018}. Separate contributions from the mini-disk regions, the accretion streams, and the circumbinary region are displayed. Figure taken from \cite{D'Ascoli-2018}.}\label{fig:f7a_dascoli18}
\end{center} 
\end{figure}

\vspace{.2cm}
To make predictions on the spectrum and the time dependence of the EM signal, \cite{D'Ascoli-2018} post-processed the data reported in \cite{Bowen-2018} with the ray-tracing code \texttt{BOTHROS} \cite[][]{Noble-2007} adopting the fast-light approximation, in which the spacetime is frozen and light propagates on the fixed background metric.
Their predictions are based on the assumption that (i) where the gas is optically thick (in the system's photosphere), it radiates a local thermal spectrum and (ii) where the gas is optically thin (mostly on the top and bottom surfaces of the disks), it radiates hard X-rays (in a manner similar to AGN). The spectra produced using \texttt{BOTHROS} can then be described in terms of two components, i.e. a thermal UV/soft X-ray portion and a coronal hard X-ray spectrum. The CBD, the accretion streams, and the mini-disks contribute to both spectral components (see Fig. \ref{fig:f7a_dascoli18}); still, the emitted power is dominated by the thermal UV (with only $\sim1\%$ radiated in the hard X-rays). Also, the majority of the flux was found to be originating from the CBD rather than from the mini-disks ($\sim$65\% and $\sim$25\%, respectively). Notably, the spectral energy distribution in Fig. \ref{fig:f7a_dascoli18} displays features consistent with predictions by the model of \cite{Tanaka-2012}, which addresses the dynamical state and thermal emission features of accretion flow onto a MBBH in a CBD.

\cite{D'Ascoli-2018} observe that, depending on the observer's viewing angle, modulations in the hard X-rays flux - with frequencies comparable to the orbital frequency - might arise due to a number of causes (i.e., Doppler beaming, gravitational lensing, periodic interaction of individual BHs with the lump).

\vspace{.3cm}

There are also other approaches that may be adopted to evolve the relativistic MHD equations over an approximate background spacetime. \cite{Mundim-2014} constructed a global, approximate metric that describes a non-spinning, equal-mass binary in quasicircular orbit by asymptotically matching different approximate solutions to Einstein's equations \cite[see also][]{Bowen-2018}. A complementary treatment was developed by \cite{Ireland-2016}, who developed an approximate spacetime which describes the inspiral of nonprecessing, spinning BBHs by matching BH perturbation theory to PN formalism.
More recently, \cite{Lopez-Armengol-2021} introduced a new approach for evolving binaries of spinning BHs, building the approximate metric as a linear superposition of two boosted Kerr-Schild BHs. This approximate spacetime (called superimposed Kerr-Schild, or SKS) was employed to produce long-term GRMHD circumbinary simulations of aligned-spinning binary black holes orbiting at fixed separation, in which the CBD is initialized by a Fishbone-Moncrief solution \cite[][]{FishBone-Moncrief-1976} and evolved solving the GRMHD equations over the background SKS metric. This system is then evolved for 266 binary periods (corresponding to approximately $t = 150\times10^3 M$), allowing for the investigation of the role of spin on the circumbinary accretion and luminosity. Specifically, spins anti-aligned (aligned) with the orbital angular momentum of the binary result in an enhanced (reduced) mass accretion rate with respect to the nonspinning case.

\vspace{.2cm}
The approximate metrics described above need the inner cavity region hosting the two MBHs to be excised. The work by \cite{Combi-2021} presented an approximate spacetime metric valid at every space point, including the BH horizons: \cite{Combi-2021} developed an analytical, time-dependent metric describing the spacetime of spinning BBHs,  constructed by superposing two Kerr metrics in harmonic coordinates; the trajectories of each BH are then described solving the PN equations of motion. The resulting metric (called superposed harmonic PN, or SHPN) is valid at every point of the computational grid, hence dispensing with the need for artificial sink terms or excision.
The SHPN metric was adopted by \cite{Combi-2022} in a GRMHD evolution of a mini-disk circumbinary accretion onto aligned-spinning BHs. The results of this simulation provided insight on the role that spin plays in the accretion rate, inflow time, and periodicities, as well as in the structure of the mini-disks.
Data from the spinning MBBHs simulation performed by \cite{Combi-2022} and from the nonspinning configurations by \cite{Bowen-2018, Bowen-2019} have been recently post-processed by \cite{Gutierrez-2022}, who extended the work done by \cite{D'Ascoli-2018} to much longer simulations. The longer duration allowed the system to relax and, consequently, to identify periodic features in the EM emission after the system has reached a quasi-steady state. These periodicities were found to be associated with the lump's dynamics, and were predicted to be present at various wavelengths. 

\vspace{.2cm}
These recent investigations pose major progress in the study of more realistic accretion flows onto massive black hole binaries, as MBHs are expected to have nonzero spin. In fact, there is observational evidence suggesting that MBHs have grown primarily by efficient accretion \citep{Marconi2004} and are expected to acquire a non-vanishing spin depending on whether accretion is prograde or retrograde, coherent or chaotic, as discussed extensively in the literature \cite[see, e.g., Refs. ][]{Gammie-2004, King2005,Berti-Volonteri-spin-2008,Dotti-spin02013,Sesana2014, Izquierdo-Villalba-2021}. As MBBHs approach merger, spins can have a major impact on the evolution of spacetime: they can alter the orbital motion, induce precession and nutation, tilt the orbital orientation and flip their sign \cite[see, e.g.,][]{Campanelli-2006b, Campanelli-2007, Healy-Lousto-2018, Gangardt-2021}. They have a key role in the mass accretion onto BHs and BBHs \cite[][]{Krolik-2005}. Also, they are an essential ingredient for the production of relativistic outflows \cite[][]{BZ-1977}.

Investigating the role of spin in determining the gas dynamics and EM emission during the late-inspiral and merger stages of massive black hole binary evolution requires a fully general relativistic treatment of the spacetime evolution.
In the following section, we give an overview of GRMHD simulations of circumbinary accretion onto MBBHs that employ fully general relativistic calculations for the evolution of the spacetime metric.

\subsubsection*{Full general relativistic MHD simulations}
The studies discussed in the previous section have produced GRMHD simulations of circumbinary accretions flows onto binaries that are evolved by Post-Newtonian gravitation, often excluding the inner region in the proximity of the BHs  and stopping the evolution at a binary separation where the PN approximation becomes inaccurate \cite[e.g., the RunIn configuration by][evolves the binary down to $r=8M$]{Noble-2012}.
To evolve a black hole binary beyond this regime, one needs to carry out fully non-linear general relativistic simulations, using numerical relativity techniques to evolve the spacetime metric by solving Einstein's equations with no approximation. This is the case of the innovative work carried out by \cite{Farris-2011, Farris-2012}, which presented full general relativistic hydrodynamic (GRHD) and GRMHD simulations of equal-mass, nonspinning binary black hole mergers in a circumbinary disk.
These simulations evolved both the pre-decoupling and post-decoupling phases of a binary-disk system.

Throughout the pre-decoupling stage (or “early-inspiral" epoch), the inspiral time scale is much longer than the orbital time scale. This fact can be exploited by neglecting the shrinking of the binary radius; to this extent, the binary is initialized adopting conformal thin-sandwich (CTS) initial conditions for a binary in quasicircular orbit and evolving the MHD equations in a metric that is quasistationary in the rotating frame of the binary \cite[analogously to the RunSE simulation by][]{Noble-2012}. Such metric is evolved employing CTS lapse and shift functions  via a simple coordinate rotation, allowing for an accurate solution of the evolving spacetime without the computational burden of a full evolution of the Einstein's equations.
This first stage of the binary evolution allows for the relaxation of the CBD and provides realistic initial data, employed to start the evolution of the late-inspiral. 
The disk is initialized by an equilibrium solution for a stationary disk around a single Kerr BH \cite[][]{Chakrabarti-1985}.
Having allowed for the relaxation of the disk to a quasistationary state, the simulation of late-inspiral and merger stages is then produced by solving the Einstein's field equations with the BSSNOK formalism with moving puncture gauge conditions \cite[][]{Campanelli-2006, Baker-2006}. 

\vspace{.2cm}
The work by \cite{Farris-2012} presented GRMHD simulations of circumbinary accretion onto BBHs that account both for the dynamical spacetime and the BH horizons and laid the groundwork for subsequent studies.
In 2014, the numerical investigation presented in \cite{Gold-2014a, Gold-2014b} extended this work by producing full GRMHD simulations of circumbinary accretion onto unequal-mass BBHs (with mass-ratio $q \equiv m_2/m_1 \in [0.1, 1]$, where $m_2 < m_1$.). This suite of simulations considered both the pre-decoupling phase \cite[][]{Gold-2014a} and the late-inspiral and merger phase \cite[][]{Gold-2014b}. Solving the full set of Einstein's field equations along with the GRMHD equations allowed \cite{Gold-2014b} to explore purely relativistic features of MBBHs, such as the formation and evolution of jets resulting from twisting the magnetic field lines anchored in the plasma, thus highlighting the importance of full GR in discovering phenomena that may be entirely missed in Newtonian and Post-Newtonian studies. The subsequent investigation by \cite{Khan-2018} tested the sensitivity of the GRMHD features of circumbinary accretion flows to the initial disk model, carrying out simulations of unequal-mass ($q=29/36$) binaries in magnetized accretion disk and surveying disk configurations characterized by different scale heights, physical extent, and magnetic-to-gas-pressure ratios. This work has shown that the general relativistic features of magnetized accretion discovered in \cite{Gold-2014a, Gold-2014b} are robust and little sensitive to the choice of the initial disk model.

Another feature of circumbinary accretion in which GR plays a key role is the \textit{formation and stability of mini-disks}. The simulations by \cite{Bowen-2018, Bowen-2019} are initialized with mini-disks (set up by quasi-hydrostationary torii around each BH) in addition to a CBD, but do not report steady-state mini-disks (see previous section). 
\cite{Gold-2014b} suggested that mini-disks could form  whenever there are stable circular orbits within the Hill sphere\footnote{The Hill spheres are regions of the spacetime where gravity is dominated by each of the binary components and are defined by the Hill radius $r_{\mathrm{Hill}}^{\pm} = 0.5\left(M^{\pm}/3M^{\mp}\right)^{1/3}a$, where $M^{\pm}$ is the mass of the primary (secondary) BH, and $a$ is the orbital separation.} around each BH. This hypothesis was tested by \cite{Paschalidis-2021}, who produced full GRMHD simulations of accreting binaries of equal-mass, spinning black holes and focused on the role of spin in the formation and stability of mini-disks. 
This work evolved four binary configurations with black holes either non-spinning or with spin aligned (antialigned) with the orbital angular momentum. To study the formation and evolution of mini-disks, the authors track the rest-mass accretion rates, the density profiles, and the mass contained within the Hill spheres of each BH. They observe that black holes with spin aligned (antialigned) to the orbital angular momentum contain more (less) mass within the Hill sphere and accrete less (more) matter. This finding is consistent with the expectation that the size of the innermost-stable-circular-orbit (ISCO) radius has a significant impact on the ability of a black hole to maintain mass within the Hill radius and form mini-disks\footnote{In fact, prograde (retrograde) orbits around spinning black holes have smaller (larger) ISCO radius.}. 
\begin{figure}
\includegraphics[width=.495\textwidth]{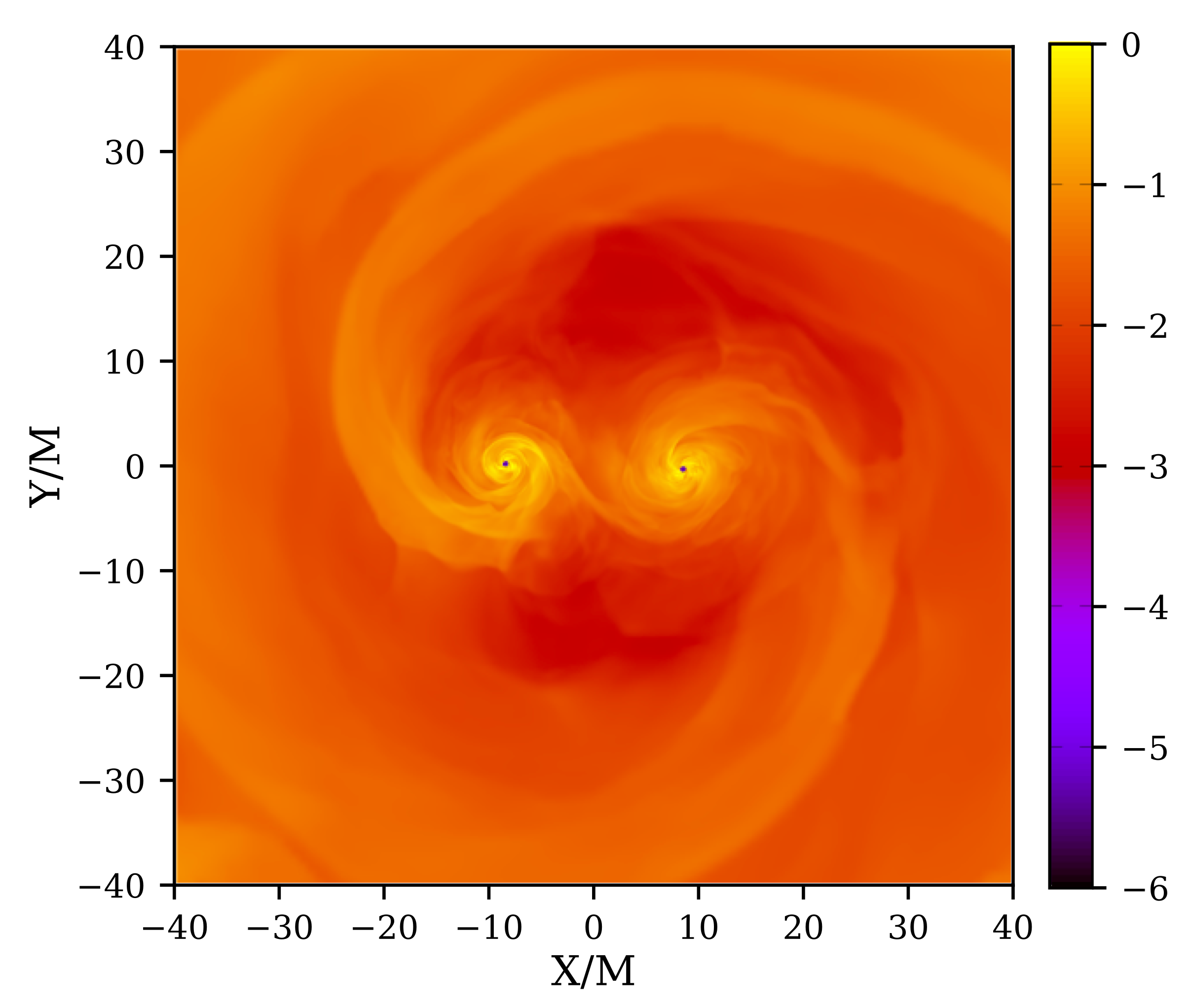}\\
\caption{Rest mass density profile on the equatorial ($xy$) plane of the aligned-spin model $\chi_{++}$ of \cite{Bright-2023} illustrating the
mini-disk structures around each black hole after  $\sim$13 binary orbits. Figure taken from \cite{Bright-2023}.}\label{fig:upup-bright}
\end{figure}

The accretion streams from the CBD form mini-disks whenever there are stable circular orbits around each black hole’s Hill sphere; thus, persistent mini-disks are expected to be present for large radial separations between the BHs, corresponding to Hill radii larger than the ISCO radius. As the binary shrinks due to the emission of GWs, the separation between the BHs will reach a threshold at $r_{Hill} \sim r_{\textrm{ISCO}}$ beyond which no stable mini-disk is able to exist (as the Hill radius decreases linearly with the binary separation). Since the spin of a BH influences the radius of the ISCO, spin is found to play a significant role in whether mini-disks can form at relativistic orbital separations. This work also has important implications for future GRMHD simulations of circumbinary systems, as it establishes that the initial binary separation should respect the criterion $r_{Hill} > r_{\textrm{ISCO}}$ to allow for the formation of mini-disks.
In addition to the mini-disk dynamics, \cite{Paschalidis-2021} also found that the Poynting luminosity associated with the collimated jet outflows launched by the binary systems is significantly larger when spinning BHs are involved.

The simulations presented in \cite{Paschalidis-2021} were extended in \cite{Bright-2023}, in which the structure of the mini-disks is examined in greater detail. We display in Fig. \ref{fig:upup-bright} the rest mass density in the equatorial plane in the aligned-spin model $\chi_{++}$ from the simulations by \cite{Bright-2023}, which exhibits clear persistent mini-disks around each of the black holes. This work also explored the periodic nature of the accretion rate and mass contained within the Hill spheres around each black hole, and found that the presence of persistent mini-disks is not only associated with lower accretion rates, but also with a suppression in the strength of the quasiperiodic modulation in the accretion rate. This result suggests that binaries at large separations may have dampened modulations, and it is not clear whether they will be able to exhibit any observable periodicity. Thus, quasiperiodic behavior in the observed EM signal arising from modulations in the mass accretion rate may not be a smoking gun for the existence of a binary. 
\begin{figure*}
\begin{center} 
\includegraphics[width=\textwidth]{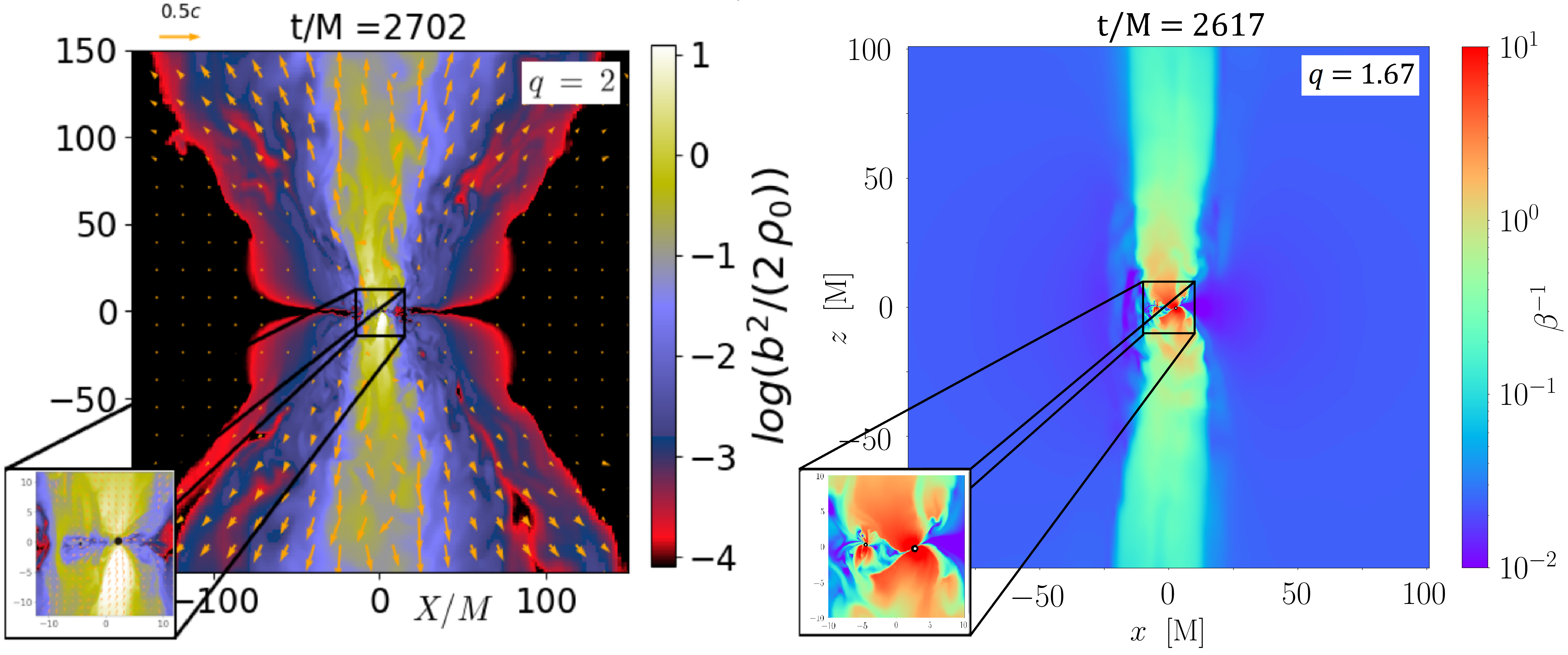}
\end{center}
\caption{Left: Magnetic-to-fluid energy density ratio parameter $\zeta^{-1} = b^2/2\rho_0$ (log scale) on a meridional plane for the $q=2$ model by \cite{Ruiz-2023}. The arrows denote the fluid velocities. Right: magnetic-to-fluid pressure ratio $\beta^{-1}$ on the meridional $xz$ plane for the \texttt{b0p6} model by \cite{Cattorini-2023}.}\label{fig:jets_conf}
\end{figure*}
The recent numerical investigation by \cite{Ruiz-2023} has provided a comprehensive exploration of the parameter space characterizing circumbinary accretion presenting full GRMHD simulations of magnetized accretion disks onto unequal-mass binaries of aligned- and misaligned-spinning MBBHs. This work investigates the impact of the binary mass-ratio and the black hole spins on the EM signatures that may emerge from the system. Considering different spin configurations (which may lie either along the initial orbital plane or an angle $\theta = \pi/4$ above it) allows to explore the impact of spin on the direction of the jet-like emission that is launched from the individual inspiraling black holes and the remnant. The results of \cite{Ruiz-2023} show that the reorientation of the remnant's spin axis drives a sudden change of the magnetic-field-lines orientation, which, however, is dampened in $\Delta t \leq 15 \ M$. This result indicates that spin-realignment at merger is not sufficient to induce helical distortions of the EM jet produced by spinning black holes.
The production and orientation of jets has also been studied in the recent work by \cite{Cattorini-2023}, which presents GRMHD calculations of merging unequal-mass binaries of misaligned-spinning black holes embedded in clouds of magnetized matter. This study shows that jets are oriented along the direction of the black hole spin axis (both across the inspiral and after merger) for distances $\leq 5 \ M$; at larger separations, the jets lose memory of the spin directions and are orientated along the $z$-axis, corresponding to the initial orientation of the magnetic field.
In Fig. \ref{fig:jets_conf}, we show a comparison between the general relativistic studies by \cite{Ruiz-2023} and \cite{Cattorini-2023}, displaying two-dimensional (meridional) slices of the magnetic-to-fluid energy density (or pressure) parameter for unequal-mass, misaligned-spinning binary configurations. 

Despite considering different environments, the results presented in these studies are consistent.

\vspace{.2cm}
\subsection{Gas cloud scenario}\label{subsec:cloud}
The balance of heating and cooling in the accretion flow in the vicinity of a MBBH can significantly alter towards the merger, when the gas is expected to be shock-heated and permeated by energetic radiation.
If radiative cooling of the accretion flow around the MBBH is inefficient, it would resemble a hot and tenuous \textit{radiatively inefficient accretion flow} \cite[RIAF][]{Ichimaru-1977, Narayan-1994}.
A RIAF tends to be hot, optically thin and geometrically thick ($h/r \gtrsim 1$), and less luminous than its radiatively efficient counterpart. The electron and ion plasma in a RIAF can form a two-temperature flow; most of the energy generated by accretion is stored within the ion plasma as thermal energy, while the electron plasma cools more efficiently and is responsible for the features of the emitted radiation.

\vspace{.2cm}
MBBH mergers in RIAFs have been explored in simulations by different groups.
General relativistic hydrodynamic (GRHD) simulations of RIAF-like accretion flows were carried out independently by \cite{Bode-2010} and \cite{Farris-2010}.
These works presented a suite of full GR simulations of equal-mass, nonspinning binary black hole mergers immersed in a gaseous environment. They found that because of high thermal velocities, the radial inflow speeds of gas are comparable to the orbital speed at a given radius. Thus, in a hot accretion flow, binary torques are not capable of clearing a central cavity because the gas ejected by the binary is  refilled on a dynamical timescale.

Aiming at identifying possible characteristic variability of the emitted EM signals, these works track the evolution of the rest-mass accretion rate, as well as the EM luminosity due to bremsstrahlung emission. \cite{Farris-2010} also considers synchrotron emission assuming the presence of a small-scale, turbulent magnetic field with a magnetic-to-gas-pressure ratio $\beta^{-1}=0.1$.
Both works established that the merger of equal-mass, nonspinning binaries in hot accretion flows features an EM flare followed by a sudden drop-off in luminosity. Additionally, \cite{Bode-2010} argued that, depending on the relative orientation between the observer and the binary, the variation in the beaming pattern of the binary could potentially give rise to modulations in the detected luminosity resulting in an “EM chirp" analogous to the gravitational one.
The subsequent follow-up by \cite{Bode-2012} extended those early results investigating the effects of varying the magnitude and orientation of the individual BH spins and the binary mass-ratio. Binaries with $q=0.5$ were found to exhibit lower and narrower luminosity peaks relative to the equal-mass case; also, $q=0.5$ binary black holes with spins misaligned with respect to the orbital angular momentum were found to produce peaks at even lower values. These results are interpreted as the consequence of the lower efficiency by unequal-mass/precessing binary torques at driving shocks in the gas.

\vspace{.2cm}
The early GRHD studies of MBBHs in gas clouds neglected the important role that magnetic fields are likely to play in driving the plasma dynamics, in the radiation emission mechanisms, and in forming relativistic jets. EM fields were first included in general relativistic electrovacuum simulations by \cite{Palenzuela-2009, Moesta-2010} and in general relativistic force-free electrodynamics (GRFFE) simulations by \cite{Palenzuela-2010, Palenzuela-2010b, Moesta-2012, Alic-2012}. These ground-breaking studies explored binary black hole mergers in a magnetically dominated plasma. Such a scenario is illustrative of conditions that could arise if a massive binary decouples from the CBD, but continues to interact with the magnetic field anchored to it. Noticeably, these simulations present evidence for the formation of separate, strongly collimated jets around each BH during the inspiral; at merger, these two jets would coalesce into a single jet around the remnant Kerr BH. The post-merger behavior of this EM outflow is well represented by the Blandford-Znajek process \cite[][]{BZ-1977}.

\begin{figure*}
\includegraphics[width=.9\textwidth]{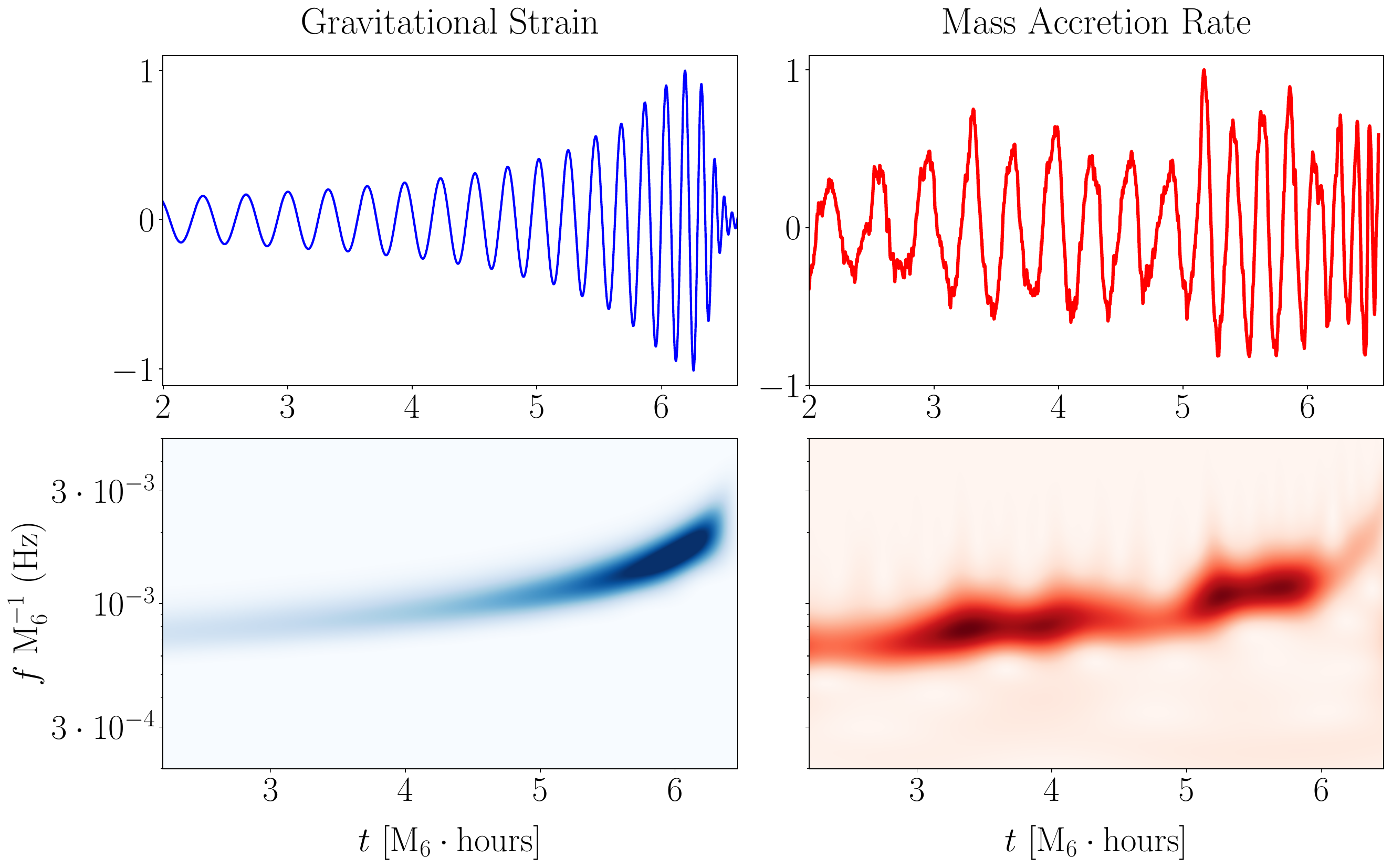}
\caption{(Top row, left): GW strain of the misaligned-spinning model \texttt{UUMIS} by \cite{Cattorini-2022} normalized to its maximum value at merger. (Top row, right): time-dependent premerger mass accretion rate over both BH horizons for \texttt{UUMIS} run, also normalized to its maximum value.
(Bottom row, left): time-frequency representation of the GW strain via wavelet PSD, showing the frequency increase of the signal over time (the so-called ``chirp''). (Bottom row, right): time-frequency representation of the accretion rate via wavelet PSD, showing similarity with the GW strain in the frequency increase over time. Time and frequency units are normalized to a binary system with total mass $M=10^6 \mathrm{M}_{\odot}$, and $\mathrm{M}_6 \equiv M/10^6 \ \mathrm{M}_{\odot}$. Figure taken from \cite{Cattorini-2022}.}
\label{fig:wavelets}
\end{figure*}

\vspace{.2cm}
More recent studies explored the behavior of \textit{moderately magnetized accretion flows} (MMAF) around binary black holes in an ideal-GRMHD context.
\cite{Giacomazzo-2012} presented ideal-GRMHD simulations of merging equal-mass, nonspinning binary black holes immersed in a MMAF using the \texttt{WhiskyMHD} code \cite[][]{Giacomazzo-2007}. 
\cite{Giacomazzo-2012} considered two models for the gas cloud surrounding the binary, both with an initially uniform rest-mass density $\rho_0$: an unmagnetized plasma, and a plasma threaded by an initially uniform magnetic field with an initial ratio of magnetic-to-fluid pressure $\beta^{-1}$ equal to 0.025. Though this study was limited to evolve a limited number of orbits ($\sim$3) before merger, it showed a rapid amplification of the magnetic field of approximately two orders of magnitude. This results showed that MMAFs exhibit different dynamics compared to unmagnetized accretion flows. The accretion flow in magnetized environments yields turbulent motion in the gas near the inspiralling BHs, eventually leading to the formation of a thin, disk-like structure rotating on the equatorial plane of the remnant Kerr BH.
Also, MMAFs can lead to strong, collimated EM emission: in fact, the Poynting luminosity resulting from their magnetized binary configuration was estimated to be approximately four orders of magnitude larger than the luminosity obtained in the GRFFE simulations by \cite{Palenzuela-2010, Moesta-2012}. This indicates that the dynamics of the late-inspiral and merger of BBHs has a strong role in driving the magnetic fields in its surroundings.

\vspace{.2cm}
The results of \cite{Giacomazzo-2012} were extended by \cite{Kelly-2017}, who covered a broader collection of physical scenarios adopting the \texttt{IllinoisGRMHD} code \cite[][]{Noble-2006, Etienne-2015} to solve the GRMHD equations. The simulations of \cite{Kelly-2017} consider equal-mass binaries of nonspinning black holes with initial separations larger than those considered in \cite{Giacomazzo-2012}. Evolving higher-separation binaries allowed the plasma in the strong-field regions to establish a quasi-equilibrium flow with the binary motion; in addition, it allows for better resolution of the timing features of the EM  emission. Several configurations differing only in the initial magnetic field value were evolved, showing that the level of the Poynting luminosity reached during the inspiral and merger is little sensitive to the initial magnetic field strength.
Also, \cite{Kelly-2017} carried out a simplified calculation of the direct EM emission generated from the plasma by post-processing the GRMHD data with the Monte Carlo radiation transport code \texttt{PANDURATA} \cite[][]{Schnittman-2013c}. \cite{Kelly-2017} assumed a simplified emission model of thermal synchrotron and found that the locally generated EM luminosity remain steady throughout the inspiral, exhibits a small burst before merger, and decreases by $\sim$50\% afterwards. Based on this simple emission model, the direct emission luminosity is orders of magnitude lower than the Poynting luminosity.

\vspace{.2cm}
The astrophysical scenario examined by \cite{Giacomazzo-2012, Kelly-2017} was further investigated in \cite{Cattorini-2021, Cattorini-2022}, which have produced three suites of GRMHD simulations of MBBHs immersed in a gas cloud and expanded the parameter space by considering binaries of spinning BHs with general spin orientation and unequal-mass ratios.
\cite{Cattorini-2021} evolve a suite of nine simulations covering a range of initially uniform, moderately magnetized fluids with different initial magnetic-to-gas-pressure ratio $\beta^{-1}_0$. For each value of $\beta_0^{-1}$, distinct spin configurations defined by adimensional spin parameters $a=(0, \ 0.3, \ 0.6)$ are considered.
For a given initial magnetization, it is shown that (aligned) spins of the individual BHs have a suppressing effect on the accretion rate as large as $\sim$48\%. In addition, spin is found to affect the peak luminosity reached shortly after merger, which is enhanced by up to a factor of $\sim$2.5 for binaries of spinning BHs compared to the nonspinning models. This intensification does not depend on the initial level of magnetization $\beta_0^{-1}$ of the gas.

The follow-up study by \cite{Cattorini-2022} evolved five equal-mass binary configurations characterized by differing spin-alignment, and investigate how the spin inclination modifies the magnetohydrodynamical behavior of the plasma during the binary late-inspiral and merger.
\cite{Cattorini-2022} identify quasiperiodic modulations in the mass accretion rate, which appear to be related to the spin-alignment. To explore the variability of these timing features, a frequency analysis is carried out calculating the power spectral density of the accretion rate for the five configurations. 
Furthermore, they perform wavelet power spectral density (WPSD) analysis on one misaligned-spinning simulation and identify a  ``chirp'' in the accretion rate that strongly resembles the gravitational one (Fig. \ref{fig:wavelets}).

The link between spin-alignment and quasiperiodicity in the accretion rate has been further investigated in the recent work by \cite{Cattorini-2023}, which produced several unequal-mass models and considered a broader family of spin misalignment and found a relation between the orientation of the individual spins relative to the orbital angular momentum and the amplitude of the quasiperiodic oscillations of the accretion rate. Also, they calculated the spin-precession parameter $\chi_{\mathrm{p}}$ \cite[][]{Schmidt-2015, Gerosa-2021b} for their misaligned-spinning models and found linear trend connecting binary systems with larger $\chi_{\mathrm{p}}$ (and hence, higher precession) to larger amplitude of the modulations in the accretion rate. This finding could provide a step toward posing an additional means to characterize spin precession and determine it observationally.

\section{Discussion}
As one of the loudest sources of GWs, merging MBBH systems provide a unique dynamical probe of strong-field general relativity and a fertile ground for the observation of fundamental physics. In the next decade, these systems will be a major target of the Laser Interferometer Space Antenna \cite[LISA,][]{LISA, LISA-2017, LISA2022}, which will detect MBBHs in the (rest-frame) mass range $\sim~[10^4-10^7] \ \mathrm{M}_{\odot}$ at rates between a few and about a hundred per year \cite[][]{Klein-2016}. Their gravitational signal is expected to be accompanied by an EM counterpart, which can be triggered by the accretion of gas onto the black holes across the inspiral, merger, and ringdown. To maximize the scientific return of LISA, several ground- and space-based missions will complement GW detection in the EM domain, looking at various frequencies for spectral and timing features characteristic of a MBBH system.
The detection of an EM counterpart would allow for an accurate identification of the GW source sky position (i.e., of the host galaxy), enabling the estimation of the source redshift either spectroscopically or photometrically.
However, the multimessenger detection of the gravitational and EM radiation emitted by a LISA source will be no easy task, since LISA sources will be localized within a sky-area ranging from hundreds of deg$^2$ to fractions of a deg$^2$ \cite[][]{Mangiagli-2020}, depending on intrinsic features of the binary and its luminosity distance.
To be univocally identified as the counterpart to a GW signal, the EM counterpart must (i) be bright enough to emit radiation above the detection limit of the EM observatory and (ii) have distinctive characteristics that would allow to discriminate among the other sources present in the observatory's field of view (e.g., the periodicities that may arise due to the dynamical coupling of the binary with its surroundings).

The features of such an emission largely depend on the amount of gas in the vicinity of the binary and on the geometry of the accretion flow.
To date, the specification of the counterpart signal to a LISA MBBH source is still unclear, due both to the scarcity of observational evidence \cite[][]{DeRosa-2020} and the complexity of producing accurate numerical simulations that take into account all the intricate physical mechanism characterizing these systems.  
To this aim, over the last years several numerical investigations of MBBHs in astrophysical environment have contributed to the theoretical characterization of these sources and advanced our understanding of the properties of the material accreting onto massive binaries, providing a fundamental step towards the accurate modeling of EM counterparts to strong LISA sources.
As a massive binary separation decreases to scales comparable to the gravitational radius, gravity becomes the dominant force in driving the fate of the system. To produce accurate numerical evolution of the late-stage of MBBH inspiral and merger it is essential to solving the equations of general relativity and magnetohydrodynamics. 

\vspace{.2cm}
As the evolution of a MBBH system is driven by emission of gravitational radiation, the morphology of the accretion flows will likely depend on the balance of heating and cooling mechanisms in the gas. Two scenarios have efficiently been investigated by GRMHD simulations, i.e., the circumbinary disk, and the gas cloud models.
Exploring different possibilities related to the binary environment allows for a deeper understanding of the physical mechanisms that can be associated to detectable EM signals.
At present, theoretical efforts involving MBBHs in circumbinary disks have incorporated relativistic magnetohydrodynamics in the numerical evolution of the accreting gas and evolved the binary spacetime with a hybrid (PN-evolved) metric or by evolving the full set of general relativistic field equations. These studies adopted the ideal GRMHD limit to probe EM signals generated from magnetized, circumbinary accretion flows onto binaries of equal/unequal-mass, nonspinning/spinning BBHs.

The timing features of the accretion rate are found to display diverse quasiperiodic behavior depending on the binary environment.
Full general relativistic explorations of circumbinary accretion observed that the accretion rate displays modulations with $f\sim0.75f_{\mathrm{orb}}$.
Conversely, circumbinary GRMHD simulations with PN gravity have shown that the accretion of mass onto the individual MBHs strongly depends on the beat frequency set by the orbital motion of the lump.

In contrast with numerical simulations of circumbinary accretion, GRMHD studies of MBBHs in magnetized gas clouds have focused on structureless uniform distributions of plasma threaded by an asymptotically uniform magnetic field, which are distinctive of hot and radiation dominated accretion flows. These studied have displayed that magnetic fields can be distorted and significantly increase their strength, developing EM jets in the polar regions above inspiraling MBHs and ultimately producing a jet around the spin axis of the remnant BH. The geometry of these structures in the vicinity of the black hole horizons depends on the system's mass ratio and the individual spins of the BHs.
Quasiperiodic oscillations of the mass accretion rate with a dominant frequency $f\sim2f_{\mathrm{orb}}$ are observed, suggesting that such modulations (which may be associated with observable EM signals) are not exclusive of environments in which a binary is embedded in a circumbinary accretion disk. Noticeably, these quasiperiodicities are linked to the orientation of the individual spins; a higher spin-misalignment corresponds to sharper quasiperiodicities in the accretion rate and to larger modulation amplitudes.

Exploring general relativistic features of diverse geometries is key to develop a coherent picture of astrophysical accretion flows onto MBBHs.
The ultimate goal of their numerical exploration is to provide accurate predictions of their electromagnetic signatures. The strongest and most valuable predictions would be a little sensitive to different details that may characterize a binary environment. In this regard, a robust validation of our results would arise from observing similar features emerging when considering different gaseous configurations \cite[see, e.g.,][]{Fedrigo-2023}.
A greater understanding of the fueling rate, geometry, and magnetic-field properties of the accreting gas will be essential in advancing the future of numerical relativity simulations and enhancing the accuracy of EM predictions. Additionally, it will be crucial to effectively incorporate various spatial scales to adequately study the evolution of the accreting gas from its early inspiral phase to the stage of merger. GRMHD simulations should also consider radiation processes in order to accurately predict EM light curves and spectra. Furthermore, they must account for different modes of accretion and incorporate radiation feedback \cite[see, e.g.,][]{Sadowski-2017}. Thus, the development of reliable radiation-transport methods in dynamical spacetime (general-relativistic radiative magnetohydrodynamics, GR-RMHD) is of utmost importance.


\bibliographystyle{aasjournal}
\bibliography{cattorini2023}

\providecommand{\noopsort}[1]{}\providecommand{\singleletter}[1]{#1}%
\begin{thebibliography}{}
\expandafter\ifx\csname natexlab\endcsname\relax\def\natexlab#1{#1}\fi
\providecommand{\url}[1]{\href{#1}{#1}}
\providecommand{\dodoi}[1]{doi:~\href{http://doi.org/#1}{\nolinkurl{#1}}}
\providecommand{\doeprint}[1]{\href{http://ascl.net/#1}{\nolinkurl{http://ascl.net/#1}}}
\providecommand{\doarXiv}[1]{\href{https://arxiv.org/abs/#1}{\nolinkurl{https://arxiv.org/abs/#1}}}

\bibitem[{Abbott {et~al.}(2016)}]{Abbott-2016}
Abbott, B.~P., {et~al.} 2016, Phys. Rev. Lett., 116, 061102,
  \dodoi{10.1103/PhysRevLett.116.061102}

\bibitem[{Abbott {et~al.}(2017)}]{Abbott-2017}
---. 2017, The Astrophysical Journal, 848, L13,
  \dodoi{10.3847/2041-8213/aa920c}

\bibitem[{Abbott {et~al.}(2021{\natexlab{a}})}]{GWTC-2.1}
---. 2021{\natexlab{a}}, arXiv e-prints, arXiv:2108.01045,
  \dodoi{10.48550/arXiv.2108.01045}

\bibitem[{Abbott {et~al.}(2021{\natexlab{b}})}]{GWTC-3}
---. 2021{\natexlab{b}}, arXiv e-prints, arXiv:2111.03606,
  \dodoi{10.48550/arXiv.2111.03606}

\bibitem[{Abbott {et~al.}(2021{\natexlab{c}})}]{Abbott-2021}
---. 2021{\natexlab{c}}, \apjl, 915, L5, \dodoi{10.3847/2041-8213/ac082e}

\bibitem[{Abrahams {et~al.}(1997)Abrahams, Anderson, Choquet-Bruhat, \&
  Jr}]{Abrahams-1997}
Abrahams, A., Anderson, A., Choquet-Bruhat, Y., \& Jr, J. W.~Y. 1997, Classical
  and Quantum Gravity, 14, A9, \dodoi{10.1088/0264-9381/14/1A/002}

\bibitem[{Alcubierre {et~al.}(2003)}]{Alcubierre-2003}
Alcubierre, M., {et~al.} 2003, Physical Reviews D, 67, 084023,
  \dodoi{10.1103/PhysRevD.67.084023}

\bibitem[{Alic {et~al.}(2012)Alic, M\"{o}sta, Rezzolla, Zanotti, \&
  Jaramillo}]{Alic-2012}
Alic, D., M\"{o}sta, P., Rezzolla, L., Zanotti, O., \& Jaramillo, J.~L. 2012,
  The Astrophysical Journal, 754, 36, \dodoi{10.1088/0004-637x/754/1/36}

\bibitem[{Amaro-Seoane {et~al.}(2012)}]{LISA}
Amaro-Seoane, P., {et~al.} 2012, arXiv e-prints, arXiv:201.3621v1.
\newblock \doarXiv{201.3621v1}

\bibitem[{Amaro-Seoane {et~al.}(2017)}]{LISA-2017}
---. 2017, arXiv e-prints, arXiv:1702.00786.
\newblock \doarXiv{1702.00786}

\bibitem[{{Amaro-Seoane} {et~al.}(2022)}]{LISA2022}
{Amaro-Seoane}, P., {et~al.} 2022, arXiv e-prints, arXiv:2203.06016,
  \dodoi{10.48550/arXiv.2203.06016}

\bibitem[{Arnowitt {et~al.}(1962)Arnowitt, Deser, \& Misner}]{Arnowitt-1962}
Arnowitt, R., Deser, S., \& Misner, C.~W. 1962, General Relativity and
  Gravitation, 40, 1997, \dodoi{10.1007/s10714-008-0661-1}

\bibitem[{Baker {et~al.}(2006)Baker, Centrella, Choi, Koppitz, \& van
  Meter}]{Baker-2006}
Baker, J.~G., Centrella, J., Choi, D.-I., Koppitz, M., \& van Meter, J. 2006,
  Phys. Rev. Lett., 96, 111102, \dodoi{10.1103/PhysRevLett.96.111102}

\bibitem[{Baumgarte \& Shapiro(1998)}]{Baumgarte-Shapiro-1998}
Baumgarte, T.~W., \& Shapiro, S.~L. 1998, Physical Reviews D, 59, 024007,
  \dodoi{10.1103/PhysRevD.59.024007}

\bibitem[{Begelman {et~al.}(1980)Begelman, Blandford, \& Rees}]{Begelman-1980}
Begelman, M.~C., Blandford, R.~D., \& Rees, M.~J. 1980, Nature, 287, 307,
  \dodoi{10.1038/287307a0}

\bibitem[{Berti {et~al.}(2007)Berti, Cardoso, Gonzalez, Sperhake, Hannam, Husa,
  \& Br\"ugmann}]{Berti-2007-unequalmass}
Berti, E., Cardoso, V., Gonzalez, J.~A., {et~al.} 2007, Phys. Rev. D, 76,
  064034, \dodoi{10.1103/PhysRevD.76.064034}

\bibitem[{{Berti} \& {Volonteri}(2008)}]{Berti-Volonteri-spin-2008}
{Berti}, E., \& {Volonteri}, M. 2008, The Astrophysical Journal, 684, 822,
  \dodoi{10.1086/590379}

\bibitem[{{Blandford} \& {Znajek}(1977)}]{BZ-1977}
{Blandford}, R.~D., \& {Znajek}, R.~L. 1977, Mon. Not. R. Astron. Soc., 179,
  433, \dodoi{10.1093/mnras/179.3.433}

\bibitem[{Bode {et~al.}(2012)Bode, Bogdanovi{\'{c}}, Haas, Healy, Laguna, \&
  Shoemaker}]{Bode-2012}
Bode, T., Bogdanovi{\'{c}}, T., Haas, R., {et~al.} 2012, The Astrophysical
  Journal, 744, 45, \dodoi{10.1088/0004-637x/744/1/45}

\bibitem[{Bode {et~al.}(2010)Bode, Haas, Bogdanovi{\'{c}}, Laguna, \&
  Shoemaker}]{Bode-2010}
Bode, T., Haas, R., Bogdanovi{\'{c}}, T., Laguna, P., \& Shoemaker, D. 2010,
  The Astrophysical Journal, 715, 1117, \dodoi{10.1088/0004-637x/715/2/1117}

\bibitem[{{Bogdanovi{\'c}} {et~al.}(2011){Bogdanovi{\'c}}, {Bode}, {Haas},
  {Laguna}, \& {Shoemaker}}]{Bogdanovic-2011}
{Bogdanovi{\'c}}, T., {Bode}, T., {Haas}, R., {Laguna}, P., \& {Shoemaker}, D.
  2011, Classical and Quantum Gravity, 28, 094020,
  \dodoi{10.1088/0264-9381/28/9/09402010.48550/arXiv.1010.2496}

\bibitem[{{Bogdanovi{\'c}} {et~al.}(2022){Bogdanovi{\'c}}, {Miller}, \&
  {Blecha}}]{Bogdanovic-2022-LR}
{Bogdanovi{\'c}}, T., {Miller}, M.~C., \& {Blecha}, L. 2022, Living Reviews in
  Relativity, 25, 3, \dodoi{10.1007/s41114-022-00037-8}

\bibitem[{Bona \& Mass\'o(1992)}]{Bona-Masso-1992}
Bona, C., \& Mass\'o, J. 1992, Phys. Rev. Lett., 68, 1097,
  \dodoi{10.1103/PhysRevLett.68.1097}

\bibitem[{Bona {et~al.}(1995)}]{Bona-1995}
Bona, C., {et~al.} 1995, Phys. Rev. Lett., 75, 600,
  \dodoi{10.1103/PhysRevLett.75.600}

\bibitem[{Bowen {et~al.}(2018)Bowen, Mewes, Campanelli, Noble, Krolik, \&
  Zilh{\~{a}}o}]{Bowen-2018}
Bowen, D.~B., Mewes, V., Campanelli, M., {et~al.} 2018, The Astrophysical
  Journal, 853, L17, \dodoi{10.3847/2041-8213/aaa756}

\bibitem[{Bowen {et~al.}(2019)Bowen, Mewes, Noble, Avara, Campanelli, \&
  Krolik}]{Bowen-2019}
Bowen, D.~B., Mewes, V., Noble, S.~C., {et~al.} 2019, The Astrophysical
  Journal, 879, 76, \dodoi{10.3847/1538-4357/ab2453}

\bibitem[{Brandt \& Br\"ugmann(1997)}]{Brandt-1997a}
Brandt, S., \& Br\"ugmann, B. 1997, Phys. Rev. Lett., 78, 3606,
  \dodoi{10.1103/PhysRevLett.78.3606}

\bibitem[{Bright \& Paschalidis(2023)}]{Bright-2023}
Bright, J.~C., \& Paschalidis, V. 2023, Monthly Notices of the Royal
  Astronomical Society, \dodoi{10.1093/mnras/stad091}

\bibitem[{{Br{\"u}gmann}(1999)}]{Bruegmann-1999}
{Br{\"u}gmann}, B. 1999, International Journal of Modern Physics D, 8, 85,
  \dodoi{10.1142/S0218271899000080}

\bibitem[{Campanelli {et~al.}(2006{\natexlab{a}})Campanelli, Lousto,
  Marronetti, \& Zlochower}]{Campanelli-2006}
Campanelli, M., Lousto, C.~O., Marronetti, P., \& Zlochower, Y.
  2006{\natexlab{a}}, Phys. Rev. Lett., 96, 111101,
  \dodoi{10.1103/PhysRevLett.96.111101}

\bibitem[{Campanelli {et~al.}(2006{\natexlab{b}})Campanelli, Lousto, \&
  Zlochower}]{Campanelli-2006b}
Campanelli, M., Lousto, C.~O., \& Zlochower, Y. 2006{\natexlab{b}}, Physical
  Reviews D, 74, 084023, \dodoi{10.1103/PhysRevD.74.084023}

\bibitem[{{Campanelli} {et~al.}(2007){Campanelli}, {Lousto}, {Zlochower},
  {Krishnan}, \& {Merritt}}]{Campanelli-2007}
{Campanelli}, M., {Lousto}, C.~O., {Zlochower}, Y., {Krishnan}, B., \&
  {Merritt}, D. 2007, Physical Reviews D, 75, 064030,
  \dodoi{10.1103/PhysRevD.75.064030}

\bibitem[{{Cattorini} {et~al.}(2023){Cattorini}, {Giacomazzo}, {Colpi}, \&
  {Haardt}}]{Cattorini-2023}
{Cattorini}, F., {Giacomazzo}, B., {Colpi}, M., \& {Haardt}, F. 2023, arXiv
  e-prints

\bibitem[{{Cattorini} {et~al.}(2021){Cattorini}, {Giacomazzo}, {Haardt}, \&
  {Colpi}}]{Cattorini-2021}
{Cattorini}, F., {Giacomazzo}, B., {Haardt}, F., \& {Colpi}, M. 2021, Physical
  Reviews D, 103, 103022, \dodoi{10.1103/PhysRevD.103.103022}

\bibitem[{{Cattorini} {et~al.}(2022){Cattorini}, {Maggioni}, {Giacomazzo},
  {Haardt}, {Colpi}, \& {Covino}}]{Cattorini-2022}
{Cattorini}, F., {Maggioni}, S., {Giacomazzo}, B., {et~al.} 2022, Astrophys.
  Journ. Lett., 930, L1, \dodoi{10.3847/2041-8213/ac6755}

\bibitem[{{Chakrabarti}(1985{\natexlab{a}})}]{Chakrabarti-1985}
{Chakrabarti}, S.~K. 1985{\natexlab{a}}, The Astrophysical Journal, 288, 1,
  \dodoi{10.1086/162755}

\bibitem[{{Chakrabarti}(1985{\natexlab{b}})}]{Chakrabarti-1985-erratum}
---. 1985{\natexlab{b}}, The Astrophysical Journal, 294, 383,
  \dodoi{10.1086/163305}

\bibitem[{Combi {et~al.}(2021)Combi, Armengol, Campanelli, Ireland, Noble,
  Nakano, \& Bowen}]{Combi-2021}
Combi, L., Armengol, F. G.~L., Campanelli, M., {et~al.} 2021, Physical Reviews
  D, 104, 044041, \dodoi{10.1103/PhysRevD.104.044041}

\bibitem[{{Combi} {et~al.}(2022){Combi}, {Lopez Armengol}, {Campanelli},
  {Noble}, {Avara}, {Krolik}, \& {Bowen}}]{Combi-2022}
{Combi}, L., {Lopez Armengol}, F.~G., {Campanelli}, M., {et~al.} 2022, The
  Astrophysical Journal, 928, 187, \dodoi{10.3847/1538-4357/ac532a}

\bibitem[{D'Ascoli {et~al.}(2018)D'Ascoli, Noble, Bowen, Campanelli, Krolik, \&
  Mewes}]{D'Ascoli-2018}
D'Ascoli, S., Noble, S.~C., Bowen, D.~B., {et~al.} 2018, The Astrophysical
  Journal, 865, 140, \dodoi{10.3847/1538-4357/aad8b4}

\bibitem[{{De Rosa} {et~al.}(2019){De Rosa}, Vignali, Bogdanović, Capelo,
  Charisi, Dotti, Husemann, Lusso, Mayer, Paragi, Runnoe, Sesana, Steinborn,
  Bianchi, Colpi, {del Valle}, Frey, Gabányi, Giustini, Guainazzi, Haiman,
  {Herrera Ruiz}, Herrero-Illana, Iwasawa, Komossa, Lena, Loiseau,
  Perez-Torres, Piconcelli, \& Volonteri}]{DeRosa-2020}
{De Rosa}, A., Vignali, C., Bogdanović, T., {et~al.} 2019, New Astronomy
  Reviews, 86, 101525, \dodoi{https://doi.org/10.1016/j.newar.2020.101525}

\bibitem[{{Dirkes}(2018)}]{Dirkes-2018-GWREV}
{Dirkes}, A. 2018, International Journal of Modern Physics A, 33, 1830013,
  \dodoi{10.1142/S0217751X18300132}

\bibitem[{D'Orazio {et~al.}(2016)D'Orazio, Haiman, Duffell, MacFadyen, \&
  Farris}]{D'Orazio-2016}
D'Orazio, D.~J., Haiman, Z., Duffell, P., MacFadyen, A., \& Farris, B. 2016,
  Monthly Notices of the Royal Astronomical Society, 459, 2379,
  \dodoi{10.1093/mnras/stw792}

\bibitem[{D'Orazio {et~al.}(2013)D'Orazio, Haiman, \&
  MacFadyen}]{D'Orazio-2013}
D'Orazio, D.~J., Haiman, Z., \& MacFadyen, A. 2013, Mon. Not. R. Astron. Soc.,
  436, 2997, \dodoi{10.1093/mnras/stt1787}

\bibitem[{{Dotti} {et~al.}(2013){Dotti}, {Colpi}, {Pallini}, {Perego}, \&
  {Volonteri}}]{Dotti-spin02013}
{Dotti}, M., {Colpi}, M., {Pallini}, S., {Perego}, A., \& {Volonteri}, M. 2013,
  The Astrophysical Journal, 762, 68, \dodoi{10.1088/0004-637X/762/2/68}

\bibitem[{Dotti {et~al.}(2012)Dotti, Sesana, \& Decarli}]{Dotti-2012}
Dotti, M., Sesana, A., \& Decarli, R. 2012, Adv. Astron., 1,
  \dodoi{10.1155/2012/940568}

\bibitem[{{Eppley}(1975)}]{Eppley-1975}
{Eppley}, K.~R. 1975, PhD thesis, Princeton University, New Jersey

\bibitem[{Etienne {et~al.}(2015)Etienne, Paschalidis, Haas, M\"{o}sta, \&
  Shapiro}]{Etienne-2015}
Etienne, Z.~B., Paschalidis, V., Haas, R., M\"{o}sta, P., \& Shapiro, S.~L.
  2015, Classical Quantum Gravity, 32, 175009,
  \dodoi{10.1088/0264-9381/32/17/175009}

\bibitem[{Farris {et~al.}(2014)Farris, Duffell, MacFadyen, \&
  Haiman}]{Farris-2014a}
Farris, B.~D., Duffell, P., MacFadyen, A.~I., \& Haiman, Z. 2014, The
  Astrophysical Journal, 783, 134, \dodoi{10.1088/0004-637x/783/2/134}

\bibitem[{Farris {et~al.}(2012)Farris, Gold, Paschalidis, Etienne, \&
  Shapiro}]{Farris-2012}
Farris, B.~D., Gold, R., Paschalidis, V., Etienne, Z.~B., \& Shapiro, S.~L.
  2012, Phys. Rev. Lett., 109, 221102, \dodoi{10.1103/PhysRevLett.109.221102}

\bibitem[{Farris {et~al.}(2010)Farris, Liu, \& Shapiro}]{Farris-2010}
Farris, B.~D., Liu, Y.~T., \& Shapiro, S.~L. 2010, Physical Reviews D, 81,
  084008, \dodoi{10.1103/PhysRevD.81.084008}

\bibitem[{Farris {et~al.}(2011)Farris, Liu, \& Shapiro}]{Farris-2011}
---. 2011, Physical Reviews D, 84, 024024, \dodoi{10.1103/PhysRevD.84.024024}

\bibitem[{{Fedrigo} {et~al.}(2023){Fedrigo}, {Cattorini}, {Giacomazzo}, \&
  {Colpi}}]{Fedrigo-2023}
{Fedrigo}, G., {Cattorini}, F., {Giacomazzo}, B., \& {Colpi}, M. 2023,
  submitted to Physical Review D

\bibitem[{{Fishbone} \& {Moncrief}(1976)}]{FishBone-Moncrief-1976}
{Fishbone}, L.~G., \& {Moncrief}, V. 1976, The Astrophysical Journal, 207, 962,
  \dodoi{10.1086/154565}

\bibitem[{Gammie {et~al.}(2004)Gammie, Shapiro, \& McKinney}]{Gammie-2004}
Gammie, C.~F., Shapiro, S.~L., \& McKinney, J.~C. 2004, The Astrophysical
  Journal, 602, 312, \dodoi{10.1086/380996}

\bibitem[{{Gangardt} {et~al.}(2021){Gangardt}, {Steinle}, {Kesden}, {Gerosa},
  \& {Stoikos}}]{Gangardt-2021}
{Gangardt}, D., {Steinle}, N., {Kesden}, M., {Gerosa}, D., \& {Stoikos}, E.
  2021, Physical Reviews D, 103, 124026, \dodoi{10.1103/PhysRevD.103.124026}

\bibitem[{{Gerosa} {et~al.}(2021){Gerosa}, {Mould}, {Gangardt}, {Schmidt},
  {Pratten}, \& {Thomas}}]{Gerosa-2021b}
{Gerosa}, D., {Mould}, M., {Gangardt}, D., {et~al.} 2021, Physical Reviews D,
  103, 064067, \dodoi{10.1103/PhysRevD.103.064067}

\bibitem[{Giacomazzo {et~al.}(2012)Giacomazzo, Baker, Miller, Reynolds, \& van
  Meter}]{Giacomazzo-2012}
Giacomazzo, B., Baker, J.~G., Miller, M.~C., Reynolds, C.~S., \& van Meter,
  J.~R. 2012, The Astrophysical Journal, 752, L15,
  \dodoi{10.1088/2041-8205/752/1/l15}

\bibitem[{Giacomazzo \& Rezzolla(2007)}]{Giacomazzo-2007}
Giacomazzo, B., \& Rezzolla, L. 2007, Classical Quantum Gravity, 24, S235,
  \dodoi{10.1088/0264-9381/24/12/s16}

\bibitem[{{Gold}(2019)}]{Gold-2019}
{Gold}, R. 2019, Galaxies, 7, 63, \dodoi{10.3390/galaxies7020063}

\bibitem[{Gold {et~al.}(2014{\natexlab{a}})Gold, Paschalidis, Etienne, Shapiro,
  \& Pfeiffer}]{Gold-2014a}
Gold, R., Paschalidis, V., Etienne, Z.~B., Shapiro, S.~L., \& Pfeiffer, H.~P.
  2014{\natexlab{a}}, Physical Reviews D, 89, 064060,
  \dodoi{10.1103/PhysRevD.89.064060}

\bibitem[{Gold {et~al.}(2014{\natexlab{b}})Gold, Paschalidis, Ruiz, Shapiro,
  Etienne, \& Pfeiffer}]{Gold-2014b}
Gold, R., Paschalidis, V., Ruiz, M., {et~al.} 2014{\natexlab{b}}, Physical
  Reviews D, 90, 104030, \dodoi{10.1103/PhysRevD.90.104030}

\bibitem[{Goodale {et~al.}(2003)Goodale, Allen, Lanfermann, Mass{\'o}, Radke,
  Seidel, \& Shalf}]{Goodale:2002a}
Goodale, T., Allen, G., Lanfermann, G., {et~al.} 2003, in Vector and Parallel
  Processing -- VECPAR'2002, 5th International Conference, Lecture Notes in
  Computer Science (Berlin: Springer).
\newblock \url{http://edoc.mpg.de/3341}

\bibitem[{{Guti{\'e}rrez} {et~al.}(2022){Guti{\'e}rrez}, {Combi}, {Noble},
  {Campanelli}, {Krolik}, {L{\'o}pez Armengol}, \&
  {Garc{\'\i}a}}]{Gutierrez-2022}
{Guti{\'e}rrez}, E.~M., {Combi}, L., {Noble}, S.~C., {et~al.} 2022, The
  Astrophysical Journal, 928, 137, \dodoi{10.3847/1538-4357/ac56de}

\bibitem[{Healy \& Lousto(2018)}]{Healy-Lousto-2018}
Healy, J., \& Lousto, C.~O. 2018, Physical Reviews D, 97, 084002,
  \dodoi{10.1103/PhysRevD.97.084002}

\bibitem[{{Healy} \& {Lousto}(2022)}]{Healy-2022_4thRITcat}
{Healy}, J., \& {Lousto}, C.~O. 2022, \prd, 105, 124010,
  \dodoi{10.1103/PhysRevD.105.124010}

\bibitem[{Hinder {et~al.}(2008)Hinder, Vaishnav, Herrmann, Shoemaker, \&
  Laguna}]{Hinder-2008-eccentric}
Hinder, I., Vaishnav, B., Herrmann, F., Shoemaker, D.~M., \& Laguna, P. 2008,
  Phys. Rev. D, 77, 081502, \dodoi{10.1103/PhysRevD.77.081502}

\bibitem[{Ichimaru(1977)}]{Ichimaru-1977}
Ichimaru, S. 1977, The Astrophysical Journal, 214, 840, \dodoi{10.1086/155314}

\bibitem[{Ireland {et~al.}(2016)}]{Ireland-2016}
Ireland, B., {et~al.} 2016, Physical Reviews D, 93, 104057,
  \dodoi{10.1103/PhysRevD.93.104057}

\bibitem[{Izquierdo-Villalba {et~al.}(2020)Izquierdo-Villalba, Bonoli, Dotti,
  Sesana, Rosas-Guevara, \& Spinoso}]{Izquierdo-Villalba-2021}
Izquierdo-Villalba, D., Bonoli, S., Dotti, M., {et~al.} 2020, Mon. Not. R.
  Astron. Soc., 495, 4681, \dodoi{10.1093/mnras/staa1399}

\bibitem[{Kelly {et~al.}(2017)Kelly, Baker, Etienne, Giacomazzo, \&
  Schnittman}]{Kelly-2017}
Kelly, B.~J., Baker, J.~G., Etienne, Z.~B., Giacomazzo, B., \& Schnittman, J.
  2017, Physical Reviews D, 96, 123003, \dodoi{10.1103/PhysRevD.96.123003}

\bibitem[{{Khan} {et~al.}(2018){Khan}, {Paschalidis}, {Ruiz}, \&
  {Shapiro}}]{Khan-2018}
{Khan}, A., {Paschalidis}, V., {Ruiz}, M., \& {Shapiro}, S.~L. 2018, Physical
  Reviews D, 97, 044036, \dodoi{10.1103/PhysRevD.97.044036}

\bibitem[{{King} {et~al.}(2005){King}, {Lubow}, {Ogilvie}, \&
  {Pringle}}]{King2005}
{King}, A.~R., {Lubow}, S.~H., {Ogilvie}, G.~I., \& {Pringle}, J.~E. 2005, Mon.
  Not. R. Astron. Soc., 363, 49, \dodoi{10.1111/j.1365-2966.2005.09378.x}

\bibitem[{Klein {et~al.}(2016)}]{Klein-2016}
Klein, A., {et~al.} 2016, Physical Reviews D, 93, 024003,
  \dodoi{10.1103/PhysRevD.93.024003}

\bibitem[{Klein(2004)}]{Klein-2004}
Klein, C. 2004, Physical Reviews D, 70, 124026,
  \dodoi{10.1103/PhysRevD.70.124026}

\bibitem[{{Krolik} {et~al.}(2005){Krolik}, {Hawley}, \& {Hirose}}]{Krolik-2005}
{Krolik}, J.~H., {Hawley}, J.~F., \& {Hirose}, S. 2005, The Astrophysical
  Journal, 622, 1008, \dodoi{10.1086/427932}

\bibitem[{{Lopez Armengol} {et~al.}(2021){Lopez Armengol}, {Combi},
  {Campanelli}, {Noble}, {Krolik}, {Bowen}, {Avara}, {Mewes}, \&
  {Nakano}}]{Lopez-Armengol-2021}
{Lopez Armengol}, F.~G., {Combi}, L., {Campanelli}, M., {et~al.} 2021, \apj,
  913, 16, \dodoi{10.3847/1538-4357/abf0af}

\bibitem[{{Lousto} \& {Healy}(2020)}]{Lousto-2020PhRvL.125s1102L}
{Lousto}, C.~O., \& {Healy}, J. 2020, \prl, 125, 191102,
  \dodoi{10.1103/PhysRevLett.125.191102}

\bibitem[{{Lovelace} {et~al.}(2010){Lovelace}, {Chen}, {Cohen}, {Kaplan},
  {Keppel}, {Matthews}, {Nichols}, {Scheel}, \& {Sperhake}}]{Lovelace-2010}
{Lovelace}, G., {Chen}, Y., {Cohen}, M., {et~al.} 2010, Physical Reviews D, 82,
  064031, \dodoi{10.1103/PhysRevD.82.064031}

\bibitem[{Maggiore(2007)}]{Maggiore-Vol1}
Maggiore, M. 2007, {Gravitational Waves: Volume 1: Theory and Experiments}
  (Oxford University Press), \dodoi{10.1093/acprof:oso/9780198570745.001.0001}

\bibitem[{Mangiagli {et~al.}(2020)Mangiagli, Klein, Bonetti, Katz, Sesana,
  Volonteri, Colpi, Marsat, \& Babak}]{Mangiagli-2020}
Mangiagli, A., Klein, A., Bonetti, M., {et~al.} 2020, Phys. Rev. D, 102,
  084056, \dodoi{10.1103/PhysRevD.102.084056}

\bibitem[{{Marconi} {et~al.}(2004){Marconi}, {Risaliti}, {Gilli}, {Hunt},
  {Maiolino}, \& {Salvati}}]{Marconi2004}
{Marconi}, A., {Risaliti}, G., {Gilli}, R., {et~al.} 2004, Mon. Not. R. Astron.
  Soc., 351, 169, \dodoi{10.1111/j.1365-2966.2004.07765.x}

\bibitem[{M\"{o}sta {et~al.}(2012)M\"{o}sta, Alic, Rezzolla, Zanotti, \&
  Palenzuela}]{Moesta-2012}
M\"{o}sta, P., Alic, D., Rezzolla, L., Zanotti, O., \& Palenzuela, C. 2012, The
  Astrophysical Journal, 749, L32, \dodoi{10.1088/2041-8205/749/2/l32}

\bibitem[{M\"osta {et~al.}(2010)M\"osta, Palenzuela, Rezzolla, Lehner, Yoshida,
  \& Pollney}]{Moesta-2010}
M\"osta, P., Palenzuela, C., Rezzolla, L., {et~al.} 2010, Physical Reviews D,
  81, 064017, \dodoi{10.1103/PhysRevD.81.064017}

\bibitem[{Mundim {et~al.}(2014)}]{Mundim-2014}
Mundim, B.~C., {et~al.} 2014, Physical Reviews D, 89, 084008,
  \dodoi{10.1103/PhysRevD.89.084008}

\bibitem[{Nakamura {et~al.}(1987)Nakamura, Oohara, \& Kojima}]{Nakamura-1987}
Nakamura, T., Oohara, K., \& Kojima, Y. 1987, Prog. Theor. Phys. Supp., 90, 1,
  \dodoi{10.1143/PTPS.90.1}

\bibitem[{{Narayan} \& {Yi}(1994)}]{Narayan-1994}
{Narayan}, R., \& {Yi}, I. 1994, The Astrophysical Journal Letters, 428, L13,
  \dodoi{10.1086/187381}

\bibitem[{Noble {et~al.}(2006)Noble, Gammie, McKinney, \& Zanna}]{Noble-2006}
Noble, S.~C., Gammie, C.~F., McKinney, J.~C., \& Zanna, L.~D. 2006, The
  Astrophysical Journal, 641, 626, \dodoi{10.1086/500349}

\bibitem[{{Noble} {et~al.}(2021){Noble}, {Krolik}, {Campanelli}, {Zlochower},
  {Mundim}, {Nakano}, \& {Zilh{\~a}o}}]{Noble-2021}
{Noble}, S.~C., {Krolik}, J.~H., {Campanelli}, M., {et~al.} 2021, \apj, 922,
  175, \dodoi{10.3847/1538-4357/ac2229}

\bibitem[{Noble {et~al.}(2009)Noble, Krolik, \& Hawley}]{Noble-2009a}
Noble, S.~C., Krolik, J.~H., \& Hawley, J.~F. 2009, The Astrophysical Journal,
  692, 411, \dodoi{10.1088/0004-637x/692/1/411}

\bibitem[{Noble {et~al.}(2007)Noble, Leung, Gammie, \& Book}]{Noble-2007}
Noble, S.~C., Leung, P.~K., Gammie, C.~F., \& Book, L.~G. 2007, Classical
  Quantum Gravity, 24, S259, \dodoi{10.1088/0264-9381/24/12/s17}

\bibitem[{Noble {et~al.}(2012)Noble, Mundim, Nakano, Krolik, Campanelli,
  Zlochower, \& Yunes}]{Noble-2012}
Noble, S.~C., Mundim, B.~C., Nakano, H., {et~al.} 2012, The Astrophysical
  Journal, 755, 51, \dodoi{10.1088/0004-637x/755/1/51}

\bibitem[{Palenzuela {et~al.}(2010{\natexlab{a}})Palenzuela, Garrett, Lehner,
  \& Liebling}]{Palenzuela-2010b}
Palenzuela, C., Garrett, T., Lehner, L., \& Liebling, S.~L. 2010{\natexlab{a}},
  Physical Reviews D, 82, 044045, \dodoi{10.1103/PhysRevD.82.044045}

\bibitem[{Palenzuela {et~al.}(2010{\natexlab{b}})Palenzuela, Lehner, \&
  Liebling}]{Palenzuela-2010}
Palenzuela, C., Lehner, L., \& Liebling, S.~L. 2010{\natexlab{b}}, Science,
  329, 927, \dodoi{10.1126/science.1191766}

\bibitem[{Palenzuela {et~al.}(2009)}]{Palenzuela-2009}
Palenzuela, C., {et~al.} 2009, Phys. Rev. Lett., 103, 081101,
  \dodoi{10.1103/PhysRevLett.103.081101}

\bibitem[{{Paschalidis} {et~al.}(2021){Paschalidis}, {Bright}, {Ruiz}, \&
  {Gold}}]{Paschalidis-2021}
{Paschalidis}, V., {Bright}, J., {Ruiz}, M., \& {Gold}, R. 2021, \apjl, 910,
  L26, \dodoi{10.3847/2041-8213/abee21}

\bibitem[{Pretorius(2005)}]{Pretorius-2005}
Pretorius, F. 2005, Phys. Rev. Lett., 95, 121101,
  \dodoi{10.1103/PhysRevLett.95.121101}

\bibitem[{{Ruiz} {et~al.}(2023){Ruiz}, {Tsokaros}, \& {Shapiro}}]{Ruiz-2023}
{Ruiz}, M., {Tsokaros}, A., \& {Shapiro}, S.~L. 2023, arXiv e-prints,
  arXiv:2302.09083, \dodoi{10.48550/arXiv.2302.09083}

\bibitem[{Sadowski \& Gaspari(2017)}]{Sadowski-2017}
Sadowski, A., \& Gaspari, M. 2017, Monthly Notices of the Royal Astronomical
  Society, 468, 1398, \dodoi{10.1093/mnras/stx543}

\bibitem[{Schmidt {et~al.}(2015)Schmidt, Ohme, \& Hannam}]{Schmidt-2015}
Schmidt, P., Ohme, F., \& Hannam, M. 2015, Physical Reviews D, 91, 024043,
  \dodoi{10.1103/PhysRevD.91.024043}

\bibitem[{Schnittman(2013)}]{Schnittman-2013a}
Schnittman, J.~D. 2013, Classical and Quantum Gravity, 30, 244007,
  \dodoi{10.1088/0264-9381/30/24/244007}

\bibitem[{Schnittman \& Krolik(2013)}]{Schnittman-2013c}
Schnittman, J.~D., \& Krolik, J.~H. 2013, The Astrophysical Journal, 777, 11,
  \dodoi{10.1088/0004-637x/777/1/11}

\bibitem[{Seidel \& Suen(1992)}]{Seidel-1992}
Seidel, E., \& Suen, W.-M. 1992, Phys. Rev. Lett., 69, 1845,
  \dodoi{10.1103/PhysRevLett.69.1845}

\bibitem[{{Sesana} {et~al.}(2014){Sesana}, {Barausse}, {Dotti}, \&
  {Rossi}}]{Sesana2014}
{Sesana}, A., {Barausse}, E., {Dotti}, M., \& {Rossi}, E.~M. 2014, The
  Astrophysical Journal, 794, 104, \dodoi{10.1088/0004-637X/794/2/104}

\bibitem[{Shi \& Krolik(2015)}]{Shi-2015}
Shi, J.-M., \& Krolik, J.~H. 2015, The Astrophysical Journal, 807, 131,
  \dodoi{10.1088/0004-637x/807/2/131}

\bibitem[{Shi {et~al.}(2012)Shi, Krolik, Lubow, \& Hawley}]{Shi-2012}
Shi, J.-M., Krolik, J.~H., Lubow, S.~H., \& Hawley, J.~F. 2012, The
  Astrophysical Journal, 749, 118, \dodoi{10.1088/0004-637X/749/2/118}

\bibitem[{Shibata \& Nakamura(1995)}]{Shibata-Nakamura-1995}
Shibata, M., \& Nakamura, T. 1995, Physical Reviews D, 52, 5428,
  \dodoi{10.1103/PhysRevD.52.5428}

\bibitem[{Smarr(1977)}]{Smarr-1977}
Smarr, L. 1977, Annals of the New York Academy of Sciences, 302, 569,
  \dodoi{https://doi.org/10.1111/j.1749-6632.1977.tb37076.x}

\bibitem[{Smarr {et~al.}(1976)}]{Smarr-1976}
Smarr, L., {et~al.} 1976, Physical Reviews D, 14, 2443,
  \dodoi{10.1103/PhysRevD.14.2443}

\bibitem[{{Smarr}(1975)}]{Smarr-1975}
{Smarr}, L.~L. 1975, PhD thesis, Texas Univ., Austin.

\bibitem[{Tanaka {et~al.}(2012)}]{Tanaka-2012}
Tanaka, T., {et~al.} 2012, Mon. Not. R. Astron. Soc., 420, 705

\bibitem[{van Meter {et~al.}(2010)van Meter, Wise, Miller, Reynolds, Centrella,
  Baker, Boggs, Kelly, \& McWilliams}]{vanMeter-2010}
van Meter, J.~R., Wise, J.~H., Miller, M.~C., {et~al.} 2010, The Astrophysical
  Journal Letters, 711, L89, \dodoi{10.1088/2041-8205/711/2/L89}

\bibitem[{Villiers {et~al.}(2003)Villiers, Hawley, \&
  Krolik}]{De_Villiers-2003}
Villiers, J.-P.~D., Hawley, J.~F., \& Krolik, J.~H. 2003, The Astrophysical
  Journal, 599, 1238, \dodoi{10.1086/379509}

\bibitem[{{Zlochower} {et~al.}(2017){Zlochower}, {Healy}, {Lousto}, \&
  {Ruchlin}}]{Zlochower-2017PhRvD..96d4002Z}
{Zlochower}, Y., {Healy}, J., {Lousto}, C.~O., \& {Ruchlin}, I. 2017, \prd, 96,
  044002, \dodoi{10.1103/PhysRevD.96.044002}

\end{thebibliography}

\end{document}